\documentclass[lettersize,journal]{IEEEtran}
\usepackage{amsmath,amsfonts}
\usepackage{algorithmic}
\usepackage{algorithm}
\usepackage{array}
\usepackage[caption=false,font=normalsize,labelfont=sf,textfont=sf]{subfig}
\usepackage{textcomp}
\usepackage{stfloats}
\usepackage{url}
\usepackage{verbatim}
\usepackage{graphicx}
\usepackage{cite}
\usepackage{svg}
\usepackage{cuted}
\usepackage{amssymb}
\usepackage{algorithm}
\usepackage{algorithmic}
\usepackage{bm} 
\usepackage{amsthm,mathtools,mathrsfs}
\usepackage{braket}
\usepackage{amsmath}

\usepackage{makecell} 
\DeclareMathOperator*{\argmax}{arg\,max}
\DeclareMathOperator*{\argmin}{arg\,min}
\usepackage{tikz}
\usetikzlibrary{positioning}
\usetikzlibrary{calc} 
\usetikzlibrary{shapes.geometric} 
\usepackage{orcidlink}

\usepackage{hhline}
\usepackage{multirow}

\theoremstyle{remark}
\newtheorem{theorem}{Theorem}

\begin{document}

\title{Structured Design of Layered Sparse Unitary Space-Time Modulation for Integrated Noncoherent and Coherent MIMO Communications}

\author{Álvaro Pendás-Recondo\orcidlink{0000-0002-8029-8024}, Enrique Pendás-Recondo\orcidlink{0000-0002-0790-7372}, Rafael González Ayestarán\orcidlink{0000-0002-9775-9448}, Jesús Alberto López-Fernández\orcidlink{0000-0001-7603-9591}, and Ignacio Santamaria \orcidlink{0000-0003-0040-7436}
\thanks{This work was supported in part by Ministerio de Ciencia e Innovación and Agencia Estatal de Investigación (MICIN)/(AEI)/10.13039/501100011033 under projects PID2023-146246OB-C32 (ANT4IT), PID2024-160783OA-I00 (RARIN-6G) and PID2022-137099NB-C43 (MADDIE); and in part by the  Gobierno del Principado de Asturias under grants Asturias-Sekuens/FEDER IDE-2024-000693, and “Severo Ochoa” Program Grant PA-22-BP21-116.

Álvaro Pendás-Recondo, Rafael González Ayestarán and Jesús Alberto López-Fernández are with the Signal Theory and Communications Group, Department of Electrical Engineering, Universidad de Oviedo, 33203, Gijón, Spain (e-mail: \{pendasalvaro,  rayestaran; jelofer\}@uniovi.es). Enrique Pendás-Recondo is with the Algebra Group, Department of Mathematics, Universidad de Oviedo, 33007, Oviedo, Spain (e-mail: pendasenrique@uniovi.es). Ignacio Santamaria is with the Advanced Signal Processing Group, Department of Communications Engineering, Universidad de Cantabria, 39005 Santander, Spain (e-mail: i.santamaria@unican.es).
}}



\maketitle

\begin{abstract}

This work presents a novel structured constellation design based on unitary space-time modulation (USTM) for integrated noncoherent and coherent multiple-input multiple-output (MIMO) communications. Information is encoded into the sparse space-time matrices through two layers: a noncoherent layer that distinguishes points on the Grassmann manifold (subspaces) and a coherent layer that distinguishes among the Stiefel representatives associated with each of those points (bases for the subspaces). We consider sequential decoding of the two layers to reduce computational complexity, while explicitly identifying the matrix structures that favor this decoding strategy. The proposed constellation builds on a previous structured Grassmannian signaling design, which has the remarkable property of producing space-time matrices with a single nonzero entry per row, while introducing a novel systematic design for the second layer corresponding to the Stiefel representatives. The design of the second layer preserves the sparsity of the first layer and significantly reduces the decoding computational complexity, while ensuring reliable performance under sequential decoding and enabling a simple yet effective bit labeling, as validated by numerical results in terms of bit error rate (BER) and symbol error rate (SER). Numerical results also show graceful performance degradation under channel correlation and, for limited coherence intervals where channel state information at the receiver (CSIR) can be estimated only imperfectly, improved performance compared with classical space-time block coding (STBC) schemes based on QAM and PSK.

\end{abstract}


\section{Introduction}
\IEEEPARstart{T}{he} acquisition of accurate channel state information (CSI) remains a critical challenge for reliable and efficient multiple-input multiple-output (MIMO) communications. At the receiver, CSI (CSIR) is generally estimated from pilot symbols transmitted at regular intervals. At the transmitter, CSI (CSIT) is typically obtained over the reverse link, either through feedback reporting the channel estimation by the receiver or, when channel reciprocity can be exploited, through pilots sent by the receiver. 
For time-varying and dispersive channels, CSIR accuracy depends on the temporal and frequency spacing of pilots, as well as the number of antennas. Pilot design thus involves an inherent trade-off between channel estimation accuracy and the reduction in spectral efficiency due to pilot overhead. CSIT acquisition is widely recognized as an even more challenging task \cite{csi1}, \cite{csi2}. In feedback-based approaches, its accuracy is limited by finite feedback rates and quantization errors, while feedback delays can leave the transmitter with outdated CSIT. Reciprocity-based approaches are also affected by channel aging, together with imperfect calibration \cite{csit_imp}.

In this context, noncoherent communication systems \cite{nonch_review}, in which neither the transmitter nor the receiver relies on CSI, have regained attention in recent years. On the one hand, the increasing operating frequencies and number of antennas considered in modern and future wireless standards further increase the challenges associated with CSI acquisition. On the other hand, noncoherent schemes are appealing for emerging wireless applications and paradigms, including ultra-reliable low-latency communications (URLLC), the Internet of Things (IoT), burst communications, and vehicular communications, including uncrewed aerial vehicles (UAVs), particularly for control and safety-critical links. Strategies for noncoherent communication systems include differential schemes \cite{diff_survey}, index modulation \cite{index_mod}, energy-based schemes \cite{energy_based}, and unitary space-time modulation (USTM), which is the focus of this work.

\subsection{Unitary Space-Time Modulation}\label{in_ustm}

USTM \cite{gt2} refers to the transmission of semi-unitary space-time matrices, i.e., matrices with orthonormal columns, or, from the perspective of differential geometry, Stiefel representatives of points in the Grassmann manifold \cite{gt3}. This modulation has been extensively studied from an information-theoretic perspective, particularly under the Rayleigh block-fading model \cite{gt2}, \cite{gt3}, \cite{gt1}, \cite{gt2_2}, in which the channel matrix is assumed to remain constant over a coherence interval before changing to an independent realization. For noncoherent detection, USTM is optimal under the assumption of equal-energy signals in this scenario, in the sense that it minimizes the union bound on the error probability \cite{gt2_2}. This property has motivated the study of constellation designs based on this scheme, also known as Grassmannian signaling, e.g., \cite{expmap2}, \cite{cubesplit}, and \cite{ub}.

Remarkably, USTM has also been identified as an appealing solution from both mutual information and error perspectives for coherent detection, i.e., when CSIR is available at the receiver \cite{gt2}, \cite{gt2_2}. However, in the coherent case, many space-time block coding (STBC) schemes are based on the transmission of multiple QAM or PSK symbols over a time block, followed by signal processing at the receiver to remove inter-symbol interference \cite{book_stbc}. The most classic example of such a scheme is Alamouti coding for two transmit antennas \cite{alamouti}. This choice is largely motivated by reduced decoding complexity, straightforward constellation design, and simple rate adaptation, consistent with the general advantages that motivate the widespread use of QAM and PSK constellations.

\subsection{Coherent versus Noncoherent Systems}\label{in_cons}

At high SNR, noncoherent MIMO can achieve a substantial fraction of the coherent capacity gain, with this fraction increasing as the coherence interval increases \cite{gt1}, \cite{gt3}. On the one hand, it is clear that for a sufficiently long coherent interval, the channel can be estimated reliably enough without significantly impacting spectral efficiency, thus favoring the coherent approach. On the other hand, for short or moderate coherence intervals, CSI is unlikely to be perfect in practical environments, making an exact information-theoretic comparison between USTM and pilot-based schemes difficult. For instance, in \cite[Fig. 8]{gt3}, the noncoherent capacity is lower than the capacity achieved with perfect CSIR but exceeds the lower bound associated with a pilot-based scheme, which is used as a reference because the exact mutual information of the latter is difficult to compute.

These considerations motivate a practical comparison, from both constellation design and error performance perspectives, of systems operating under limited coherence intervals. In this context, USTM offers significant potential to efficiently integrate noncoherent and coherent communications using the same constellation and signal structure, enabling a system that exploits both noncoherent and coherent detection for maximum efficiency across a wide range of wireless scenarios. This proposal is further supported by the analysis in \cite{gt3}, which shows that, for limited coherence intervals, the insights on the suitable number of antennas for pilot-based schemes align with those for the noncoherent case, facilitating efficient integration of both architectures into a common hardware system.

Such an integrated approach has been previously proposed, e.g., in \cite{2non}, \cite{multicasting}, and \cite{multires2}, in all cases by encoding information into one noncoherent layer, or up to two in \cite{2non}, together with a single coherent layer. In \cite{multires2}, Grassmannian signaling is used for noncoherent communication, while coherent detection enables additional information to be conveyed through the indices of the active transmit antennas. Particularly related to this work is \cite{multicasting}, where the coherent layer is encoded through the specific Stiefel representatives selected to transmit the Grassmannian points carrying the noncoherent information. Nevertheless, \cite{multicasting} considers a multicasting setting in which some users can access only the noncoherent layer, whereas others can recover both the noncoherent and coherent information.

Finally, although addressing a different application, another approach related to the integration of noncoherent and coherent schemes is the use of Grassmannian signaling to convey additional information through the pilot symbols of a coherent system \cite{p1}, \cite{p2}.

\subsection{Contributions}\label{in_cont}
The main contributions of this work are summarized as follows:

\begin{itemize}
    \item We present a novel USTM constellation design for integrated noncoherent and coherent MIMO communications. Following the idea proposed in \cite{multicasting}, although here we focus on a single-user scenario and present a novel and different constellation design, the information is organized into two layers: a noncoherent layer that distinguishes points on the Grassmann manifold and a coherent layer that distinguishes among the Stiefel representatives associated with each of those points. Detection of both layers is performed sequentially to reduce computational complexity, which then scales with the sum of the constellation sizes of the individual layers rather than their product, as in joint coherence detection. While the diversity order (or asymptotic SNR exponent) of this decoder is analyzed in \cite{multicasting}, we provide a complementary analysis of the matrix structures that favor symbol error rate (SER) performance under sequential decoding. Specifically, we identify constellation properties that allow the second layer to be correctly detected even when detection of the first layer fails.
    
    \item The proposed constellation builds on the structured design proposed in \cite{gd} for the first layer (points in the Grassmann manifold) while presenting a novel design for the second layer (Stiefel representatives). The design in \cite{gd} is structured in the sense that it provides deterministic solutions with the notable property that the resulting space-time matrices contain a single nonzero element per row, and can therefore be regarded as an sparse design \cite{sparse}, which provides potential advantages for hardware implementation. We propose a novel second-layer design that preserves this sparsity and its associated advantages while ensuring reliable error performance under sequential decoding with significantly reduced computational complexity. In addition, it enables simple yet effective bit labeling for enhanced bit error rate (BER) performance.
    
    \item Numerical evaluations based on Monte Carlo simulations of SER and BER under Rayleigh block-fading show the scalability of the proposed design with respect to the number of antennas and transmission rate (constellation size), without compromising error performance under sequential decoding compared with alternative USTM strategies. Additionally, the results show that the error performance degrades gracefully under channel correlation. Finally, we compare the proposed solution with STBC schemes based on QAM or PSK and purely noncoherent Grassmannian signaling in what is arguably the most relevant scenario for its application: limited coherence intervals, where CSIR is imperfect due to the limited number of available pilot symbols. In this regime, the proposed layered USTM scheme outperforms STBC schemes while requiring significantly lower computational complexity than fully noncoherent Grassmannian schemes that forgo channel estimation. When CSIR cannot be obtained, as may occur for even shorter coherence intervals, the proposed design naturally reduces to Grassmannian signaling, which can be regarded as a particular case of the proposed framework. To the best of our knowledge, this is the first practical comparison of these strategies in terms of SER and BER performance under imperfect CSIR.

    \item Finally, we relate the observed SER and BER results to previous information-theoretic findings in the literature, showing that USTM signaling, when the constellations of both layers are properly designed, constitutes an appealing strategy from capacity, error, and practical implementation perspectives across a wide range of MIMO scenarios without CSIT.

\end{itemize}

The remainder of this paper is organized as follows. Section \ref{PR} provides the theoretical background on Stiefel and Grassmann manifolds required for this work. Section \ref{SM} presents the system model for integrated noncoherent and coherent communications. Section \ref{ME} reviews the performance metrics that guide the constellation design proposed in Section \ref{CD}. Section \ref{NR} presents the numerical SER and BER results. Finally, conclusions are drawn in Section \ref{Conclusion}.

\textit{Notation}: Scalars are denoted by lower case or upper case italic letters, $x$, $X$, while matrices are denoted by bold uppercase letters, $\mathbf{X}$. The transpose, conjugate transpose, determinant, trace and range of a matrix are denoted by $\left(\cdot\right)^T$, $\left(\cdot\right)^H$, $\operatorname{det}\left(\cdot\right)$, $\operatorname{tr}\left(\cdot\right)$, and $\operatorname{range}(\mathbf{\cdot})$, respectively. The identity matrix of size $M$ is written as $\mathbf{I}_M$. The Frobenius norm of a matrix is written as $\|\mathbf{X}\|_F$. Subspaces and sets are denoted by calligraphic letters, $\mathcal{X}$.  The absolute value of a scalar, or the cardinality of a set, is denoted by $\left|\cdot\right|$. The real and imaginary parts of a scalar or matrix are denoted by $\mathfrak{Re}(\cdot)$ and $\mathfrak{Im}(\cdot)$, respectively. The imaginary unit is denoted as $i$. The expected value is denoted as $\operatorname{E}\{\cdot\}$. A circularly-symmetric complex Gaussian distribution with mean $\mu$ and variance $\sigma^2$ is denoted by $\mathcal{CN}\left(\mu,\sigma^{2}\right)$.

\section{Preliminaries}\label{PR}
In this section, we present the theoretical background of this work. We refer the reader to \cite{ghb}, \cite{geom_alg}, and \cite{cgrass} for further details on these notions.

\subsection{Stiefel and Grassmann manifolds}
The {\em complex Stiefel manifold} $\textup{St}_{\mathbb{C}}(T,M)$, with $T\geq M$, is defined as the set of all matrices in $\mathbb{C}^{T\times M}$ whose columns are orthonormal:
\begin{equation*}
    \textup{St}_{\mathbb{C}}(T,M) \coloneqq \left\{\mathbf{U} \in \mathbb{C}^{T\times M}: \mathbf{U}^H\mathbf{U}=\mathbf{I}_M \right\}.
\end{equation*}
Each $\mathbf{U} \in \textup{St}_{\mathbb{C}}(T,M)$ generates an $M$-dimensional complex subspace $\mathcal{U}$ of $\mathbb{C}^T$, in the sense that the columns of $\mathbf{U}$ form an orthonormal basis of $\mathcal{U}$. We write in this case $\mathcal{U}=\operatorname{span}(\mathbf{U})$. The set of all $M$-dimensional complex subspaces of $\mathbb{C}^T$ is called the {\em complex Grassmann manifold} or {\em complex Grassmannian} $\textup{Gr}_{\mathbb{C}}(T,M)$, which is given by
\begin{equation*}
    \textup{Gr}_{\mathbb{C}}(T,M) \coloneqq \left\{\operatorname{span}(\mathbf{U}): \mathbf{U} \in \textup{St}_{\mathbb{C}}(T,M) \right\}.
\end{equation*}
Although each subspace $\mathcal{U}=\operatorname{span}(\mathbf{U})$ can be represented by the matrix $\mathbf{U} \in \textup{St}_{\mathbb{C}}(T,M)$, this representation is not unique. Indeed, two orthonormal bases generate the same subspace if and only if they differ by a unitary transformation. In other words, $\mathbf{U}_1, \mathbf{U}_2 \in \textup{St}_{\mathbb{C}}(T,M)$ satisfy $\operatorname{span}(\mathbf{U}_1)=\operatorname{span}(\mathbf{U}_2)$ if and only if $\mathbf{U}_2=\mathbf{U}_1\mathbf{R}$, where $\mathbf{R}$ is an element of the {\em unitary group}
\begin{equation*}
    U(M) \coloneqq \left\{\mathbf{R} \in \mathbb{C}^{M\times M}: \mathbf{R}^H\mathbf{R}=\mathbf{I}_M=\mathbf{R}\mathbf{R}^H \right\}.
\end{equation*}
All matrices that generate the same subspace $\mathcal{U}=\operatorname{span}(\mathbf{U}) \in \textup{Gr}_{\mathbb{C}}(T,M)$ are denoted as $[\mathbf{U}]$ and called the {\em equivalence class} of $\mathbf{U}$:
\begin{equation*}
    [\mathbf{U}] \coloneqq \left\{\mathbf{U}\mathbf{R} \in \textup{St}_{\mathbb{C}}(T,M): \mathbf{R} \in U(M)\right\}.
\end{equation*}
In this way, each subspace $\mathcal{U}=\operatorname{span}(\mathbf{U})$ uniquely corresponds to $[\mathbf{U}]$, so we can make the identification $[\mathbf{U}]\equiv \mathcal{U}$ and say that $[\mathbf{U}]$ is a {\em point} in $\textup{Gr}_{\mathbb{C}}(T,M)$, whereas the {\em point} $\mathbf{U}$ in $\textup{St}_{\mathbb{C}}(T,M)$ is called a {\em Stiefel representative} of $\mathcal{U}$.

Given $[\mathbf{U}]\equiv \mathcal{U}$, the matrix $\mathbf{P} \coloneqq \mathbf{U}\mathbf{U}^H \in \mathbb{C}^{T\times T}$ is the unique {\em orthogonal projector} onto $\mathcal{U}$, i.e., it is the unique matrix that satisfies the properties $\mathbf{P}^H=\mathbf{P}$, $\mathbf{P}^2=\mathbf{P}$, and $\operatorname{range}(\mathbf{P})=\mathcal{U}$. This projector is independent of the chosen Stiefel representative.

\begin{figure}
\centering
\includegraphics[width=0.8\linewidth]{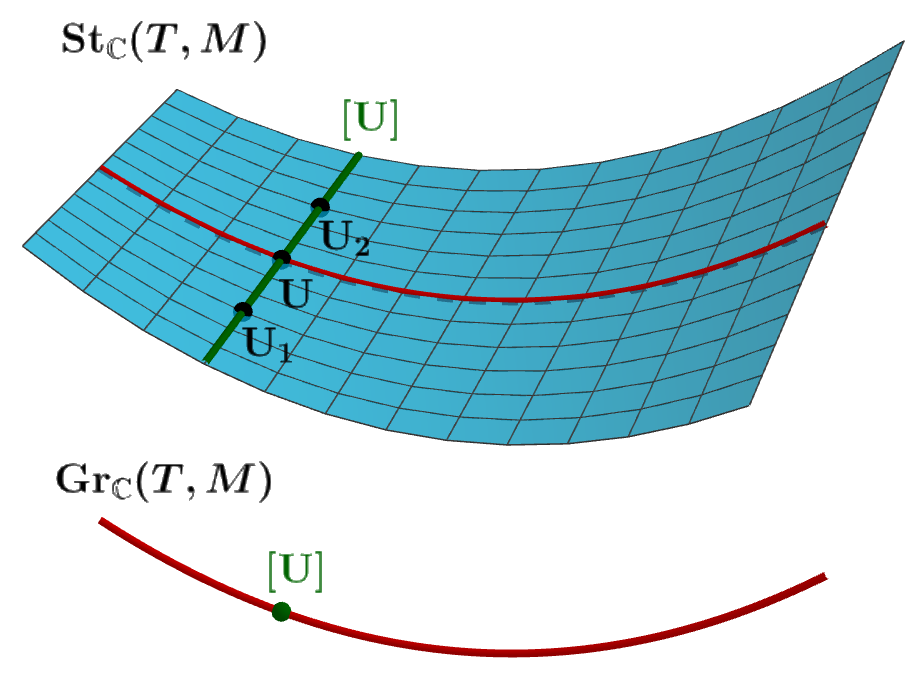}
\caption{Conceptual visualization of the interplay between the Stiefel and Grassmann manifolds. The points $\mathbf{U}, \mathbf{U}_1,\mathbf{U}_2 \in \textup{St}_{\mathbb{C}}(T,M)$ are all Stiefel representatives of the same Grassmannian point $[\mathbf{U}]\in \textup{Gr}_{\mathbb{C}}(T,M)$, i.e., $[\mathbf{U}]$ corresponds to an entire submanifold of $\textup{St}_{\mathbb{C}}(T,M)$.}
\label{fig:stiefel_grassmann}
\end{figure}

Both $\textup{St}_{\mathbb{C}}(T,M)$ and $\textup{Gr}_{\mathbb{C}}(T,M)$ are real manifolds of dimensions $2TM-M^2$ and $2M(T-M)$, respectively. In addition, each $[\mathbf{U}] \subset \textup{St}_{\mathbb{C}}(T,M)$ can be seen as an embedded submanifold of $\textup{St}_{\mathbb{C}}(T,M)$ of dimension $M^2$ (see Figure~\ref{fig:stiefel_grassmann}).

\subsection{Principal angles}
Given two points $[\mathbf{U}_1], [\mathbf{U}_2] \in \textup{Gr}_{\mathbb{C}}(T,M)$, the {\em principal angles} $\theta_1,\ldots,\theta_M \in \left[0,\frac{\pi}{2}\right]$ quantify the ``relative position'' of the subspaces $\mathcal{U}_1=\operatorname{span}(\mathbf{U}_1)$ and $\mathcal{U}_2=\operatorname{span}(\mathbf{U}_2)$. They are defined as
\begin{equation}
\label{eq:principal_angles}
        \theta_m \coloneqq \arccos{(\sigma_m)} \in \left[0,\frac{\pi}{2}\right], \quad m = 1,\ldots, M,
\end{equation}
where $\sigma_m\in [0,1]$ is the $m$-th largest singular value of $\mathbf{U}_1^H\mathbf{U}_2$. These values are independent of the chosen Stiefel representatives $\mathbf{U}_1, \mathbf{U}_2 \in \textup{St}_{\mathbb{C}}(T,M)$.

The geometric interpretation of the principal angles is the following. If one principal angle is zero, then $\mathcal{U}_1$ and $\mathcal{U}_2$ share a nontrivial direction. If $\theta_m=0$ for all $m$, then $\mathcal{U}_1 = \mathcal{U}_2$. Conversely, if $\theta_m\not=0$ for all $m$, then $\mathcal{U}_1 \cap \mathcal{U}_2 = \{0\}$, which is the most relevant situation in the present work.

On the other hand, if one principal angle is $\frac{\pi}{2}$, there is a direction in $\mathcal{U}_1$ orthogonal to $\mathcal{U}_2$, and vice versa. If $\theta_m=\frac{\pi}{2}$ for all $m$, then $\mathcal{U}_1$ and $\mathcal{U}_2$ are mutually orthogonal: every direction in $\mathcal{U}_1$ is orthogonal to every direction in $\mathcal{U}_2$.

\subsection{Vertical angles}
There is a natural analogue of principal angles for two points $\mathbf U_1,\mathbf U_2\in\mathrm{St}_{\mathbb C}(T,M)$ belonging to the same equivalence class, i.e., satisfying $[\mathbf U_1]=[\mathbf U_2]$. In this situation, all principal angles \eqref{eq:principal_angles} are equal to zero, and therefore they do not capture the remaining degrees of freedom associated with the choice of orthonormal basis. We thus seek a measure of the ``relative orientation'' between the two orthonormal frames represented by the columns of $\mathbf U_1$ and $\mathbf U_2$.

As explained above, two matrices in the same equivalence class differ by a unique unitary matrix $\mathbf{R}=\mathbf{U}_1^H\mathbf{U}_2$, which completely characterizes the difference between the two Stiefel representatives. Specifically, if a vector in the common subspace $\operatorname{span}(\mathbf{U}_1)=\operatorname{span}(\mathbf{U}_2)$ is represented by the coordinate vector $v \in \mathbb{C}^{M\times 1}$ with respect to the basis $\mathbf{U}_2$, then its coordinates with respect to the basis $\mathbf{U}_1$ are given by $\mathbf{R}v$. Since $\mathbf R \in U(M)$, all its eigenvalues lie on the unit circle and can be written as
\begin{equation}\label{spa}
        \mathrm{eig}(\mathbf{R}) = \left\{e^{i\phi_1}, e^{i\phi_2}, \ldots , e^{i\phi_M}\right\}.
\end{equation}
We call the phases $\phi_1,\ldots,\phi_M \in (-\pi,\pi]$ the {\em vertical angles} between $\mathbf U_1$ and $\mathbf U_2$. Geometrically, the magnitudes $|\phi_m|$ quantify the phase rotations induced by the unitary transformation relating the two frames. If one vertical angle is zero, then the corresponding eigenvector is left unchanged by $\mathbf R$. This means that there exists a direction in the common subspace whose coordinate representation is identical in both orthonormal frames. If $\phi_m=0$ for all $m$, then $\mathbf R=\mathbf I_M$ and consequently $\mathbf U_1=\mathbf U_2$. Conversely, if $\phi_m\not=0$ for all $m$, then $\mathbf R$ has no eigenvalue equal to $1$, i.e., there is no direction in the subspace whose coordinates are identical in both orthonormal bases. On the other hand, if $|\phi_m|=\pi$ for some $m$, the corresponding eigenvector undergoes the largest possible phase change. If $|\phi_m|=\pi$ for all $m$, then $\mathbf R=-\mathbf I_M$, so that every basis vector is reversed and $\mathbf U_2=-\mathbf U_1$.

\section{System Model}\label{SM}

\subsection{Communication System}\label{cs}
We consider a MIMO wireless communication system with $M$ transmit antennas and $N$ receive antennas over a frequency-flat block-fading channel with a coherence time of $T_{\mathrm{c}}$ symbols, where $T_{\mathrm{c}} \geq 2M$. Accordingly, the channel matrix $\mathbf{H}\in\mathbb{C}^{M\times N}$ remains constant over each coherence block of $T_{\mathrm{c}}$ symbols. Note that, depending on the scenario, frequency and bandwidth, the frequency-flat assumption may require integration with orthogonal frequency-division multiplexing (OFDM) and the insertion of a cyclic prefix. Nonetheless, in such a case, all techniques, constellations, and considerations presented in this paper remain valid, except for the feature offered by the sparse constellation designs of requiring a single active transmit antenna per time slot. This is further discussed in Section \ref{STBC}. The signal at the receiver over a time block of $T$ symbols, $T\leq T_{\mathrm{c}}$, is denoted by $\mathbf{Y} \in \mathbb{C}^{T \times N}$ and is given by 

\begin{equation}\label{Y}
\mathbf{Y} = \mathbf{X}\mathbf{H} + \sqrt{\frac{M}{T\rho}}\mathbf{W},
\end{equation} where $\mathbf{W}\in \mathbb{C}^{T \times N}$ represents the additive white Gaussian noise (AWGN), i.e., its entries are i.i.d. with $w_{ij}\sim \mathcal{C}\mathcal{N}(0,1)$ and $\rho$ denotes the signal-to-noise ratio (SNR). By default, we assume a Rayleigh block-fading channel model, i.e., the entries of $\mathbf{H}$ are also i.i.d. with $h_{ij}\sim\mathcal{CN}(0,1)$.

We propose a layered USTM system based on the transmission of space-time matrices $\mathbf{X} \in \mathbb{C}^{T \times M}$ such that $\mathbf{X}^{H}\mathbf{X} = \mathbf{I}_M$. Each matrix $\mathbf{X}$ is a Stiefel representative of a point $[\mathbf{X}]$ in the complex Grassmann manifold $\textup{Gr}_{\mathbb{C}}(T,M)$, as described in Section \ref{PR}. Communication and decoding are organized into two layers: a noncoherent layer that can be decoded without CSIR, and a coherent layer that requires CSIR, i.e., an estimation of $\mathbf{H}$ at the receiver, denoted by $\hat{\mathbf{H}}$. To this end, it is useful to define the codebook, or constellation, as 

\begin{equation}\label{alphabet}
\mathcal{X}= \mathcal{U}\mathcal{R} = \{ \mathbf{U}_g \mathbf{R}_s: g=1,2,\ldots,L_{\mathrm{G}}; s=1,2,\ldots,L_{\mathrm{S}}\},
\end{equation} such that $\left|\mathcal{X}\right| = L_\mathrm{G}L_{\mathrm{S}} = L$, where $L_\mathrm{G}$ denotes the number of constellation points in the Grassmannian, and $L_{\mathrm{S}}$ denotes the number of different Stiefel representatives for each of those points. The set $\mathcal{U}= \{ \mathbf{U}_1, \mathbf{U}_2, \ldots, \mathbf{U}_{L_\mathrm{G}}\}$ contains one explicitly selected representative for each distinct point on the Grassmannian, whereas the set $\mathcal{R}= \{ \mathbf{R}_1, \mathbf{R}_2, \ldots, \mathbf{R}_{L_\mathrm{S}}\}$ contains the unitary matrices used to generate all the different Stiefel representatives. In this manner, $\mathbf{X}_{gs} = \mathbf{U}_g\mathbf{R}_s$ denotes the transmission of the $g$-th point in the Grassmann manifold using the $s$-th Stiefel representative associated with that point. This design ensures that $\mathbf{X}_{gs}$ and $\mathbf{X}_{gk}$ represent exactly the same point in the Grassmannian (belong to the same equivalence class), which implies $\mathbf{X}_{gs}\mathbf{X}_{gs}^H = \mathbf{X}_{gk}\mathbf{X}_{gk}^H$, even when $s \neq k$ (Section \ref{PR}). Importantly, note that the principal angles between points in the Grassmannian are independent of the chosen representatives (Section \ref{PR}). Finally, assuming equiprobable symbols, the first and second layers transmit with rates $R_{\mathrm{G}} = \log_2 \left(L_{\mathrm{G}}\right)/T$ and $R_{\mathrm{S}} = \log_2 \left(L_{\mathrm{S}}\right)/T$, respectively.

\subsection{First Layer Decoding}\label{FS}
In case no CSIR is available, only the first communication layer is used by transmitting symbols in the alphabet $\mathcal{U}$, which results in a standard Grassmannian signaling scheme \cite{nonch_review}. For the noncoherent decoding of the first layer,  the optimal ML detector that minimizes error probability is  \cite{gt2}

\begin{equation}\label{mlg}
\hat{g} = \argmax_{\mathbf{U}_{g}\in\mathcal{U}} \operatorname{tr}\left( \mathbf{Y}^H \mathbf{P}_{[\mathbf{U}_{g}]}\mathbf{Y}\right) = \argmax_{\mathbf{U}_{g}\in\mathcal{U}} \|\mathbf{U}_{g}^H \mathbf{Y} \|_F^2,
\end{equation} where $\mathbf{P}_{[\mathbf{U}_{g}]} = \mathbf{U}_{g}\mathbf{U}_{g}^H$ is the projector onto the subspace $\operatorname{span}(\mathbf{U}_{g})$ (Section \ref{PR}). Importantly, the performance of this decoder is unaffected by the potential activation of the second layer, since $\mathbf{P}_{[\mathbf{U}_{g}]} = \mathbf{P}_{[\mathbf{X}_{lk}]}$ if $g=l$ (Section \ref{cs}), and the first layer is decoded correctly as long as $\hat{g} = g$, regardless of the value of $\hat{s}$ for the second one.


\subsection{Second Layer Decoding}\label{SL}
If CSIR is available due to the transmission of pilots, the second layer can be activated and transmit symbols are chosen from $\mathcal{X}$. The tradeoff between the spectral efficiency sacrificed by pilot transmission and the additional rate provided by the second layer for a given $T_{\mathrm{c}}$ is evaluated for different scenarios in Section \ref{NR}. 

In that case, the receiver has an estimation (not necessarily perfect) of the channel matrix, $\hat{\mathbf{H}}$. Assuming that $\hat{\mathbf{H}}$ is reliable enough, optimum decoding requires joint coherent detection of both layers, that is

\begin{equation}\label{mls_joint}
\left(\hat{g},\hat{s}\right) = \argmin_{\mathbf{U}_{\hat{g}}\in\mathcal{U}, \mathbf{R}_{\hat{s}}\in\mathcal{R}} \|\mathbf{Y} - \mathbf{U}_g\mathbf{R}_s \hat{\mathbf{H}} \|_F^2.
\end{equation}

However, the computational cost of this decoder, which is a major concern in USTM schemes, represents a significant practical limitation. We therefore adopt the decoding strategy proposed in \cite{multicasting}, where the index of the first layer, $\hat{g}$, is still decoded as if there were a single layer. As noted, the performance of (\ref{mlg}) is unaffected by using $\mathcal{X}$ instead of $\mathcal{U}$, since it only distinguishes between different points in the Grassmannian. With these considerations, the detector for the second layer, which is going to work with the $\hat{g}$ estimation provided by the detector for the first layer, is

\begin{equation}\label{mls}
\hat{s} = \argmin_{\mathbf{R}_{s}\in\mathcal{R}} \|\mathbf{Y} - \mathbf{U}_{\hat{g}}\mathbf{R}_s \hat{\mathbf{H}} \|_F^2.
\end{equation}

As shown in \cite{multicasting}, combining this detector with (\ref{mlg}) yields a suboptimal decoding strategy for coherent detection compared with joint decoding in (\ref{mls_joint}). However, both approaches achieve the same diversity order, with an asymptotic SNR exponent of $-MN$ as long as the diversity product (DP) of both layers is not zero, while sequential decoding provides a substantial reduction in computational complexity, as discussed in Section \ref{CC}. For the BER evaluation presented in Section \ref{NR}, we denote the BER of the first and second layers by $\mathrm{BER}_1$ and $\mathrm{BER}_2$, respectively, and define the BER of the overall system as $\mathrm{BER} = (\log_2(L_{\mathrm{G}})\mathrm{BER}_1 + \log_2(L_{\mathrm{S}})\mathrm{BER}_2)/\log_2(L)$. The same notation is used for the SER. In case both layers employ the same symbol rate, i.e., $R_\mathrm{G}=R_\mathrm{S}$, the overall SER is given by $\mathrm{SER} = (\mathrm{SER}_1 + \mathrm{SER}_2)/2$. In case $R_\mathrm{G} \neq R_\mathrm{S}$, there is no unique definition of the aggregated $\mathrm{SER}$, since symbols carry different amounts of information, unlike bits.

Note that the second layer is decoded correctly as long as $\hat{s} = s$, regardless of the value of $\hat{g}$. Intuitively, one might expect that if the Grassmannian point or subspace is decoded incorrectly in the first layer ($\hat{g} \neq g$), the second-layer decoding would almost always fail, leading to $\mathrm{SER}_2 \geq \mathrm{SER}1$. However, a remarkable feature of the design proposed in Section \ref{CD}, although not unique to it, is that, even when the first decoder selects the wrong subspace, the second decoder may still correctly identify the index of the basis used to represent that incorrect subspace with a probability significantly higher than that of a random guess.

\section{Metrics for Constellation Design}\label{ME}

In this section, we present the metrics used to guide constellation design for both communication layers, informed by prior studies in the literature. We first review the metrics related to the error performance of noncoherent communications using Grassmannian signaling (first layer). We then review the metrics associated with decoding points on the Stiefel manifold with CSIR (second layer). In both cases, Rayleigh block-fading, as described in Section \ref{SM}, is assumed.

\subsection{First Layer} \label{FLM}
The study of symbol error probability using Grassmannian signaling and the ML noncoherent decoder described in (\ref{mlg}) has been addressed in several previous works, with a summary provided in \cite[Section III]{gd}. Briefly, many studies approach the problem by maximizing the minimum pairwise chordal distance between constellation points, \cite{do1}, \cite{grassma}, which is defined as

\begin{equation}\label{dc}
\begin{aligned}
\textrm{d}_\textrm{G,c}\left( \mathbf{X}_i, \mathbf{X}_j \right) &= \frac{1}{\sqrt{2}}\| \mathbf{P}_{[\mathbf{X}_i]}-\mathbf{P}_{[\mathbf{X}_j]} \|_F \\
&= \left(\sum_{m=1}^{M}\sin^2\theta_m\right)^{1/2}.
\end{aligned}
\end{equation} 

The chordal distance has the advantage of being differentiable at every point on the manifold, which enables gradient-descent optimization \cite{do1}. Note that the analysis presented in \cite[Section V]{gt2} describes how the pairwise error probability (PEP) decreases with the singular values of $\mathbf{X}_i^H \mathbf{X}_j$, which correspond exactly to the cosines of the principal angles between both points. However, this relationship is established through a product rather than a sum, \cite[Eq. (17)]{gt2}. Consequently, the so-called DP, or coherence criterion \cite{do2}, \cite{do3}, has also been considered as a constellation design metric, being defined as

\begin{equation}\label{dp}
\begin{aligned}
\textrm{DP}_{\textrm{G}}(\mathbf{X}_i, \mathbf{X}_j) &= \textrm{det} \left(\mathbf{I}_M - \mathbf{X}_i^H\mathbf{X}_j\mathbf{X}_j^H\mathbf{X}_i  \right)\\
&= \prod_{m=1}^M \sin^2\theta_m,
\end{aligned}
\end{equation} where, as with the chordal distance, the objective is to maximize the minimum pairwise value. Of particular relevance is the study carried out in \cite{do3}, where it is shown that, at high SNR, the DP is a more suitable metric than the chordal distance, while the latter remains an important design criterion under low-SNR conditions.

Finally, other works such as \cite{ubp} and \cite{ub} consider the minimization of the asymptotic PEP union bound (UB), which is defined in \cite{ub}, at high SNR and up to a constant factor, as

\begin{equation}\label{ubc}
\textrm{UB}_{\textrm{G}}(\mathcal{X}_{\textrm{G}}) = \sum_{i<j} \textrm{det} \left(\mathbf{I}_M - \mathbf{X}_i^H\mathbf{X}_j\mathbf{X}_j^H\mathbf{X}_i  \right) ^{-N}, 
\end{equation} where all possible pairs are considered within a single metric, together with the number of receive antennas. However, when $\left(\mathbf{I}_M - \mathbf{X}_i^H \mathbf{X}_j \mathbf{X}_j^H \mathbf{X}_i\right)$ is not full rank, as commonly occurs for constellations using sparse matrices, its determinant vanishes, which makes certain error-probability expressions based on the UB inadequate for those cases \cite{sparse}.

\subsection{Second Layer} \label{SLM}

The PEP of ML decoding of two points in the Stiefel manifold $\mathbf{X}_i$ and $\mathbf{X}_j$ given CSIR was derived in \cite[Section V]{gt2} for Rayleigh block-fading. As a relevant conclusion, the PEP decreases as the singular values of $\mathbf{X}_i - \mathbf{X}_j$ increase, and that relation is also established through a product \cite[Eq. (19)]{gt2}. Furthermore, in case $\mathbf{X}_i$ and $\mathbf{X}_j$ belong to the same equivalence class, we can write $\mathbf{X}_j = \mathbf{X}_i\mathbf{R}$, where $\mathbf{R}$ is a unitary matrix, $\mathbf{R} \in U(M)$ (Section \ref{PR}). With all of these considerations, it is of interest to define a DP criterion for such Stiefeld points as:

\begin{equation}\label{dps}
\begin{aligned}
\textrm{DP}_\mathrm{S} (\mathbf{X}_i, \mathbf{X}_j) &= \textrm{det} \left( \left(\mathbf{X}_i - \mathbf{X}_j\right)^H \left(\mathbf{X}_i - \mathbf{X}_j \right) \right)\ \\ &= 4^M\prod_{m=1}^M \sin^2 \left( \frac{\phi_m}{2}\right) = \prod_{m=1}^M \sigma^2_{\mathrm{S},m},
\end{aligned}
\end{equation} where $\sigma_{\mathrm{S},m}\in [0,1]$ is the $m$-th largest singular value of $\mathbf{X}_i-\mathbf{X}_j$, and $\phi_m$ denotes the vertical angles between $\mathbf{X}_i$ and $\mathbf{X}_j$ (Section \ref{PR}).

It should be noted that \cite[Section V]{gt2} identified the criteria of maximizing the singular values of $\mathbf{X}_i^H\mathbf{X}_j$ for noncoherent detection and minimizing the singular values of $\mathbf{X}_i-\mathbf{X}_j$ for coherent detection as incompatible. However, this observation was made in the context of a unified single-step approach in which any points on the Stiefel manifold can be selected. In contrast, our work adopts a sequential approach to constellation design, as described in Section \ref{CD}. First, well-separated points on the Grassmann manifold are obtained by selecting specific Stiefel representatives. Subsequently, for each of those representatives, we seek to obtain well-separated points in the same equivalent class (i.e., within the same subspace), rather than across the entire Stiefel manifold. In a certain sense, we prioritize the first criterion over the second; however, this choice is justified because the coherent decoder inherently benefits from the availability of CSIR, as shown in Section \ref{NR}. Moreover, the conclusion of incompatibility supports our approach, as it indicates that the two metrics cannot be optimized simultaneously via a single search over the Stiefel manifold.

\section{Constellation Design}\label{CD}

We now present the proposed constellation design. For the first layer, we adopt the design introduced in \cite{gd}. We then describe a novel design for the second layer, specifically constructed to operate jointly with the first layer, resulting overall in a USTM strategy capable of operating under both coherent and noncoherent conditions. The error performance is evaluated in Section \ref{NR}.

\subsection{First Layer}\label{FCD}

The constellation for the first layer, $\mathcal{U}$, is obtained following the approach presented in \cite{gd}, whose code is publicly available. This design is based on the geodesic curves of the Grassmann manifold, exploiting its geometric structure to provide deterministic solutions with error performance comparable to optimization-based approaches. Furthermore, all matrices in $\mathcal{U}$ exhibit the remarkable property of containing a single nonzero element per row, which allows the proposed construction to be regarded as a sparse design \cite{sparse}.

On the other hand, this design is restricted to a number of points $L_{\textrm{G}} \leq 4M^2$. Therefore, our objective in the second-layer design is to preserve the single-nonzero-element property, which reduces decoding complexity (Section \ref{CC}), while providing greater flexibility in the total data rate by allowing arbitrary values of $L_{\textrm{S}}$ (and consequently of $L$). Finally, this design assumes that $T = 2M$ (which is the minimum block length for Grassmannian signaling), an assumption that we also maintain for the second layer. This system choice is further justified in \cite[Section II]{gd} and is evaluated in this work in Section \ref{NR} against larger values of $T$ when $T_{\mathrm{c}}>2M$.

\subsection{Second Layer}\label{SCD}
The first-layer design yields the alphabet $\mathcal{U}=\{\mathbf{U}_g \in \mathbb{C}^{T\times M}: g=1,\ldots,L_{\mathrm{G}}\}$. Once this codebook has been obtained, the complete constellation $\mathcal{X}$ in \eqref{alphabet} is given by matrices of the form $\mathbf{X}_{gs} = \mathbf{U}_{g} \mathbf{R}_s$, with $\mathbf R_s\in U(M)$ and $s=1,2,\ldots,L_{\mathrm{S}}$. Namely, for a fixed $g \in \{1,\ldots,L_{\mathrm{G}}\}$, each choice of a unitary matrix $\mathbf{R}_s$ provides a different Stiefel representative $\mathbf{X}_{gs} \in \textup{St}_{\mathbb{C}}(T,M)$ of the same Grassmannian point $[\mathbf{U}_g] \in \textup{Gr}_{\mathbb{C}}(T,M)$. Therefore, the goal is to find a suitable set $\mathcal{R}=\{\mathbf{R}_s: s=1,\ldots,L_{\mathrm{S}}\}$ for our purposes.

First, to preserve the single nonzero entry per row, we set
\begin{equation}
\label{eq:Rs}
    \mathbf{R}_s = \operatorname{diag}\left( e^{i\phi_{s1}}, e^{i\phi_{s2}}, \ldots, e^{i\phi_{sM}}\right),
\end{equation}
so that the vertical angles between $\mathbf{X}_{gs}$ and $\mathbf{X}_{gk}$ (Section \ref{PR}) are given by 
\begin{equation}
\label{eq:vert_angles}
    \operatorname{wrap}(\phi_{km}-\phi_{sm}) \in (-\pi,\pi], \quad m=1,\ldots,M,
\end{equation}
where $\operatorname{wrap}(x) \coloneqq \operatorname{Arg}(e^{ix})$, being $\operatorname{Arg}$ the principal argument with values in $(-\pi,\pi]$. Since only the absolute values of \eqref{eq:vert_angles} are relevant for the DP in \eqref{dps}, we define
\begin{equation}\label{alpha}
\alpha_m ^{(s,k)} \coloneqq \left| \operatorname{wrap}\left(\phi_{km} - \phi_{sm}\right)\right| \in [0, \pi],
\end{equation} and

\begin{equation}\label{Psi}
\Psi (\mathcal{R}) = \min_{s\neq k} \prod_{m=1}^{M} \sin^2 \left(\frac{\alpha_m^{(s,k)}}{2}\right),
\end{equation} with the maximization of this criterion being equivalent to maximizing the minimum pairwise DP in \eqref{dps}.

In order to maximize \eqref{Psi}, we first choose $\phi_{s1}$ as evenly spaced angles on the unit circle:
\begin{equation}
\label{eq:initial}
    \phi_{s1}=\frac{2\pi}{L_{\mathrm{S}}}s, \quad s=1,\ldots,L_{\mathrm{S}}.
\end{equation}
Next, for $\phi_{s2}$, and subsequently for $\phi_{s3},\ldots,\phi_{sM}$, instead of assigning these angles to the same indices $s$ (which would make the same points the closest in every dimension $m$), in order to maximize \eqref{Psi}, we assign the same set of angles to different indices choosing the best permutation among a predefined set of candidates.

Since an exhaustive search over all possible permutations quickly becomes computationally prohibitive as $L_{\mathrm{S}}$ increases, we apply the following algebraic result (see \cite{permu}): $p(x)=Ax+Bx^2$ is a {\em permutation polynomial} modulo $L_{\mathrm{S}}$ over the ring $\mathbb{Z}_{L_{\mathrm{S}}} \coloneqq \{ 1,2,\ldots,L_{\mathrm{S}}\equiv0\}$ if and only if $A \in \mathbb{Z}_{L_{\mathrm{S}}} $ is odd and $B \in \mathbb{Z}_{L_{\mathrm{S}}}$ is even. This means that, fixing $A$ and $B$, the values $p(1),\ldots,p(L_{\mathrm{S}})$ form a permutation of $1,\ldots,L_{\mathrm{S}}$. Consequently, we can use $p(s)$ to permute the assignment of the angles, choosing suitable values of $A$ and $B$ for each $m$. Importantly, the construction presented in \cite{permu} is the only approach that enables deterministic permutations over $L_{\mathrm{S}}$ to be generated using a polynomial, which is desirable from a computational perspective, while the number of available permutations can be readily increased by increasing the polynomial degree. The restriction to use a permutation polynomial of degree 2 is due to the fact that, if we use a polynomial of degree 1, there is very limited freedom (only one coefficient) to rearrange which pairs have small vertical angles. On the other hand, if we increase the degree to three or more coefficients, the computation of \eqref{eq:a_b} becomes substantially more expensive while providing negligible gains in the presented results.

Since $A$ must be odd and $B$ even, we set $A=2a+1$ and $B=2b$, with $a,b \in \mathcal{P} \coloneqq \{0,1,\ldots,\frac{L_{\mathrm{S}}}{2}-1\}$, so that the permutation polynomial becomes
\begin{equation}
    p_{a,b}(s)=(2a+1)s+2bs^2 \quad (\operatorname{mod}L_{\mathrm{S}})
\end{equation}
(with the convention $x \operatorname{mod}L_{\mathrm{S}} \in \{1,\ldots,L_S\}$) and then, for each diagonal element $m=1,\ldots,M$ in \eqref{eq:Rs}, we select a candidate pair $(a_m,b_m)\in \mathcal{P}\times\mathcal{P}$ that yields
\begin{equation}\label{pac_phi}
\phi_{sm} = \frac{2\pi}{L_{\mathrm{S}}} p_{a_m,b_m}(s) \in (0,2\pi],
\end{equation}
with the initial choices $a_1=b_1=0$, consistently with \eqref{eq:initial}. Note that, for $s\not=k$, since $p_{a_m,b_m}(s) \not= p_{a_m,b_m}(k)$, this ensures that $\alpha_m^{(s,k)} > 0$.

\begin{figure}
\centering
\includegraphics[width=1\linewidth]{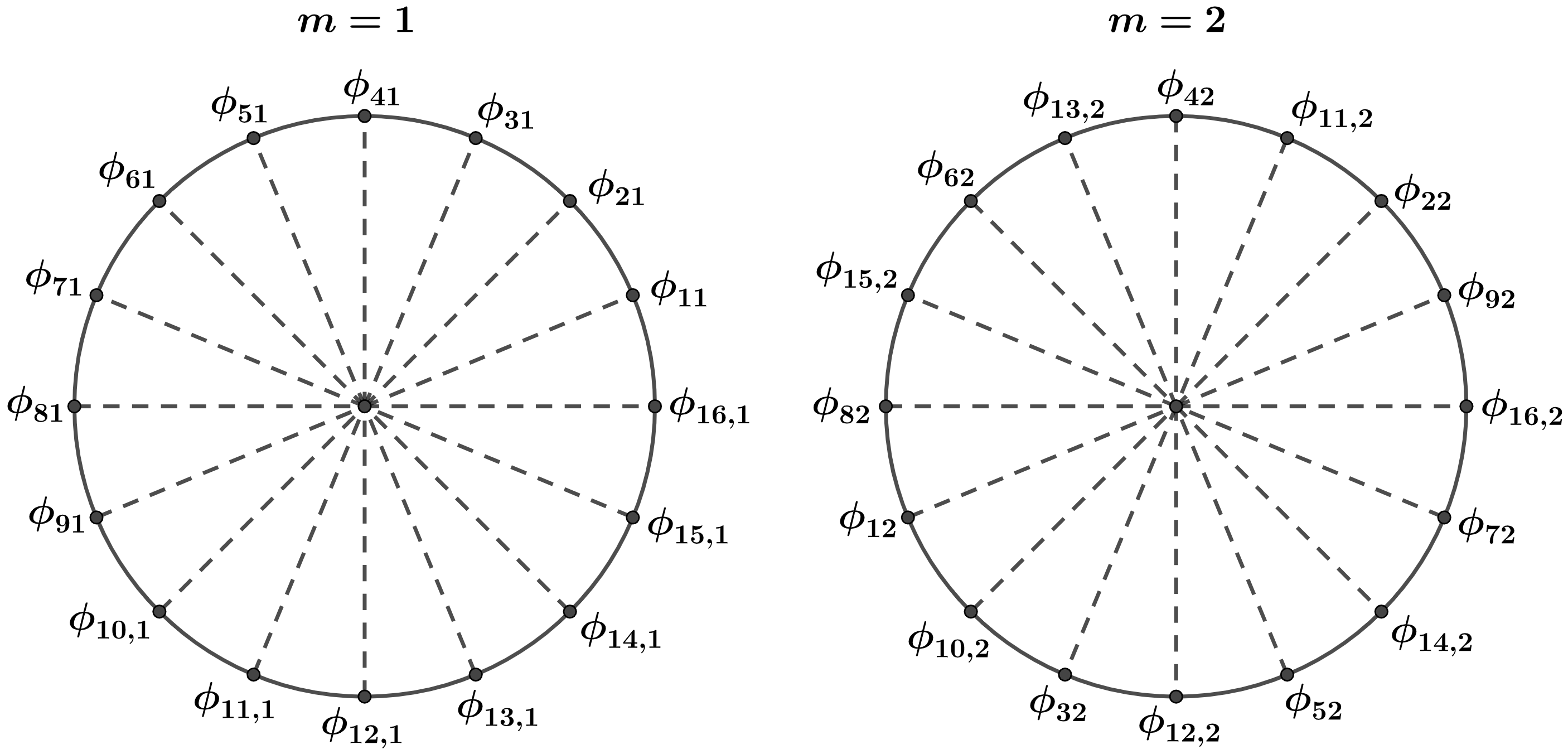}
\caption{Assignment of the angles $\phi_{sm}$ for $M=2$ and $L_{\mathrm{S}}=16$. The unit circle is divided into $L_{\mathrm{S}}$ equally spaced angles, and for $m=1$, $\phi_{s1}$ are assigned to these angles in strictly increasing order, according to \eqref{eq:initial}. For $m=2$, the computation in \eqref{eq:a_b} yields $a_2=0$ and $b_2=4$, so that \eqref{pac_phi} permutes the assignment order (e.g., $p_{a_2,b_2}(1)=9$). For $m>2$, the values $(a_m,b_m)$ keep changing, resulting in a different permutation for each $m$, with the aim of maximizing \eqref{Psi}.}
\label{fig:angles}
\end{figure}

For $m=2,\ldots,M$, we approach the selection of $(a_m,b_m)$ in a sequential and greedy manner. Identifying $\alpha_m^{(s,k)} \coloneqq \alpha_{a_m,b_m}^{(s,k)}$, we choose each $(a_m,b_m)$ in a recursive manner (starting with $a_1=b_1=0$) as
\begin{equation}
\label{eq:a_b}
(a_m,b_m)
=
\underset{
\begin{subarray}{c}
  (a,b) \in \mathcal{P}\times \mathcal{P}   
\end{subarray}
}
{\operatorname{arg\,max}}
\;
\underset{s\neq k}{\min}
\left[
\left(
\prod_{j=1}^{m-1}
\alpha_j^{(s,k)}
\right)
\alpha_{a_m,b_m}^{(s,k)}
\right],
\end{equation} with the resulting permutation illustrated in Fig. \ref{fig:angles} for the case $M=2$ and $L_{\mathrm{S}}=16$.

\subsection{Bit Labeling}\label{BL}

For bit labeling, we adopt a simple yet effective approach in which pairs of codewords with the largest Hamming distance are assigned to constellation points with the largest DP. The systematic procedure for applying this approach to the first layer is described in \cite[Section IV-C]{gd}. For the second layer proposed here, we adopt a sequential approach: the first codeword (in bits) is assigned to $\mathbf{X}_{g1}$, and the codeword with the largest Hamming distance from it is assigned to $\mathbf{X}_{gb}$, with $b=2,3,\ldots,L_{\mathrm{S}}$, such that $\mathrm{DP}_{\mathrm{S}}(\mathbf{X}_{g1},\mathbf{X}_{gb})$ is maximized. This procedure is then repeated recursively for the remaining codewords and constellation points. Note that, due to the proposed systematic design, this procedure only needs to be computed for a single value of $g$ and can then be directly adopted for the remaining values.

\subsection{Sequential Decoding Analysis}\label{SDCon}
An additional motivation for using structured constellation designs, as proposed here, follows from this mathematical result:

\begin{theorem}
\label{thm:UR}
    Let $[\mathbf{U}_g],[\mathbf{U}_{\tilde{g}}] \in \mathrm{Gr}_{\mathbb{C}}(T,M)$. If the diagonal entries of $\mathbf{U}_g^H\mathbf{U}_{\tilde{g}}$ are real and non-negative, then
    \begin{equation}
    \label{eq:proof}
        ||\mathbf{U}_{g}\mathbf{R}_{s}-\mathbf{U}_{\tilde{g}}\mathbf{R}_{s}||_F \leq ||\mathbf{U}_{g}\mathbf{R}_{s}-\mathbf{U}_{\tilde{g}}\mathbf{R}_{\tilde{s}}||_F
    \end{equation}
    for all diagonal $\mathbf{R}_{s},\mathbf{R}_{\tilde{s}} \in U(M)$. If, in addition, $\mathbf{U}_g^H\mathbf{U}_{\tilde{g}}$ is hermitian and positive semi-definite, then \eqref{eq:proof} holds for all $\mathbf{R}_{s},\mathbf{R}_{\tilde{s}} \in U(M)$.
\end{theorem}
\begin{proof}
    See Appendix\ref{app:proof}.
\end{proof}

Designing a constellation for which many points $\mathbf{X}_{gs} = \mathbf{U}_g\mathbf{R}_s \in \mathcal{X}$ satisfy \eqref{eq:proof} implies that, as anticipated in Section \ref{SL}, even when detection of the first layers fails, i.e., $\hat{g}\neq g$ in \eqref{mlg}, the second layer may still be correctly detected in many instances. Specifically, assuming $\mathbf{X}_{gs}$ is transmitted, and denoting the incorrect first-layer estimate by $\hat{g}=\tilde{g}$, the correct index for the second layer $s$ remains the closest in terms of Frobenius distance among the Stiefel representatives associated with $[\mathbf{U}_{\tilde{g}}]$ when \eqref{eq:proof} holds, and is therefore more likely to be selected in \eqref{mls}. 

Two features of the proposed constellation design are directly related to these considerations. First, since the considered matrices $\mathbf{R}_s \in U(M)$ are diagonal, \eqref{eq:proof} holds whenever the diagonal entries of $\mathbf{U}_g^H\mathbf{U}_{\tilde{g}}$ are real and nonnegative (importantly for our design, this includes zero). Second, in the context of a sparse design with a single nonzero element per row, if the nonzero entries that overlap between $\mathbf{U}_g$ and $\mathbf{U}_{\tilde{g}}$ have the same complex phase, then the diagonal entries of $\mathbf{U}_g^H\mathbf{U}_{\tilde{g}}$ are real and nonnegative (sufficient but not necessary condition). Furthermore, the first-layer design sets the upper submatrix of every $\mathbf{U}_g$ to a scaled version of $\mathbf{I}_M$ \cite{gd}, so the overlapping entries across different points corresponding to those positions always have the same phase. As a result, this matrix structure ensures that many pairs of points automatically satisfy \eqref{eq:proof}. To provide quantitative examples, for $T=4$, $M=2$, and $L_{\mathrm{G}}=16$, 84 out of the 120 possible pairs satisfy \eqref{eq:proof}, with 64 of these cases guaranteed by the sparse matrix structure. For $T=8$, $M=4$, and $L_{\mathrm{G}}=64$, 1632 out of 2016 pairs satisfy \eqref{eq:proof}, of which 1536 are directly guaranteed by the matrix structure. The resulting SER performance of sequential decoding with the proposed constellation designs, compared with unstructured alternatives, is presented in Section \ref{NR}.

\subsection{Computational Complexity}\label{CC}
We now address the decoding computational complexity, which is an important concern for Grassmannian or Stiefel signaling with ML decoding. For the first layer, the computational complexity of evaluating the Frobenious norm in (\ref{mlg}) is dominated by the matrix multiplication $\mathbf{X}^H \mathbf{Y}$, which, for arbitrary Grassmannian constellations, results in a total complexity of $\mathcal{O}(L_{\mathrm{G}}MTN)$ when comparing the $L_{\mathrm{G}}$ possible codewords. Similarly, the computational cost of the second layer in (\ref{mls}) is dominated by the matrix multiplication and results in a complexity of $\mathcal{O}(L_{\mathrm{S}}MTN)$.

If the second layer is activated, the first layer could also be decoded using CSIR, and a search over both the $g$ and $s$ indices could be performed as in (\ref{mls_joint}). However, this results in a total computational complexity of $\mathcal{O}(L_{\mathrm{G}}L_{\mathrm{S}}MTN)$, instead of $\mathcal{O}((L_{\mathrm{G}}+L_{\mathrm{S}})MTN)$, which is a major advantage of the chosen sequential decoding strategy. Moreover, the proposed constellation design further reduces these values significantly. First, since each row of $\mathbf{X}$ contains a single nonzero entry whose position is known to the receiver, all two-element matrix multiplications involving $\mathbf{X}$ or $\mathbf{X}^H$ can be implemented by simply adding scaled rows, thus computational complexity of the first layer resulting in  $\mathcal{O}(L_{\mathrm{G}}TN)$, corresponding to an improvement by a factor of $M$. Second, under sequential decoding, using $\mathbf{X}_{\hat{g}s}=\mathbf{U}_{\hat{g}}\mathbf{R}_s$,
the second-layer decoder first computes
$\mathbf{Z}=\mathbf{U}_{\hat{g}}^{H}\mathbf{Y}$ in $\mathcal{O}(TN)$ operations, since each row of $\mathbf{U}_{\hat{g}}$ has only one nonzero entry. Since $\mathbf{R}_s$ is diagonal and unitary, an equivalent decoding score uses $M$ scalar correlations
between corresponding rows of $\mathbf{Z}$ and $\hat{\mathbf{H}}$. These correlations are independent of $s$, so they are computed once in $\mathcal{O}(MN)$ operations and reused for all $L_{\mathrm{S}}$ candidate matrices. Evaluating this score for each of the $L_{\mathrm{S}}$ candidate matrices costs $\mathcal{O}(M)$ per candidate, giving a search complexity of $\mathcal{O}(L_{\mathrm{S}}M)$. Thus, the overall  complexity of the second layer is $\mathcal{O}(TN+MN+L_{\mathrm{S}}M)=\mathcal{O}(TN+L_{\mathrm{S}}M)$, since $T\geq 2M$, resulting in a total complexity of $\mathcal{O}(L_{\mathrm{G}}TN+ TN + L_{\mathrm{S}}M) = \mathcal{O}(L_{\mathrm{G}}TN + L_{\mathrm{S}}M)$. This represents an improvement by a factor of $TN$ in the second layer compared with arbitrary USTM space-time matrices.

Given this analysis, for computational purposes, it is desirable to select values of $L_{\mathrm{G}}$ and $L_{\mathrm{S}}$ such that $L_{\mathrm{G}},L_{\mathrm{S}} \ll L = L_{\mathrm{G}}L_{\mathrm{S}}$. Equivalently, values such that $R_{\mathrm{G}} \approx R_{\mathrm{S}}$, so that the decoding complexity is not dominated by either of the two layers. However, when $L_{\mathrm{G}} \neq L_{\mathrm{S}}$, choosing $L_{\mathrm{S}} > L_{\mathrm{G}}$ is more favorable from a computational complexity perspective and, provided that CSIR is sufficiently accurate, also in terms of error performance, as observed in Section \ref{NR}.

\section{Numerical Results}\label{NR}

In this section, we assess the error performance of the designed constellations through Monte Carlo simulations of the SER and BER. All results are obtained for the block-fading channel (with potential correlation) introduced in Section \ref{SM}, using by default the sequential decoding approach described in Section \ref{SM}. The SNR corresponds to the parameter $\rho$ in \eqref{Y}. For spatial correlation, we adopt the standard Kronecker model \cite{kron}, parameterized by the transmitter and receiver correlation factors $\gamma_{\mathrm{t}},\gamma_{\mathrm{r}} \in [0,1]$, respectively, where $0$ denotes uncorrelated antenna channels and $1$ denotes fully correlated channels, with all antennas experiencing identical fading and thus providing no spatial diversity.

\subsection{SER Results}\label{SERR}

\begin{figure*}
\centering
\subfloat[\normalsize $R_{\mathrm{G}}=1$,  $R_{\mathrm{S}}=1$ $(\textrm{bps/Hz})$ \label{fr1_1}]{%
      \includegraphics[width=0.33\linewidth]{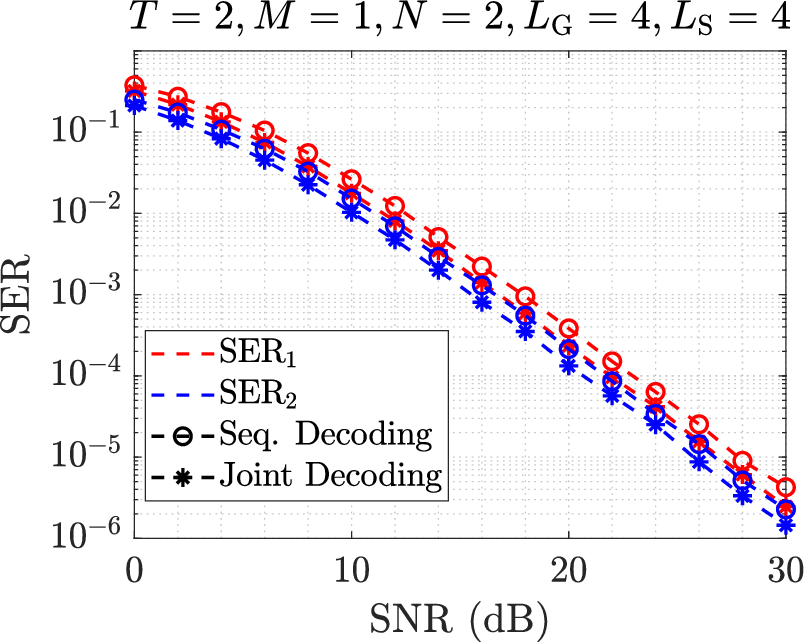}}
      \hfill
\subfloat[\normalsize  $R_{\mathrm{G}}=1$,  $R_{\mathrm{S}}=1$ $(\textrm{bps/Hz})$   \label{fr1_2}]{%
      \includegraphics[width=0.33\linewidth]{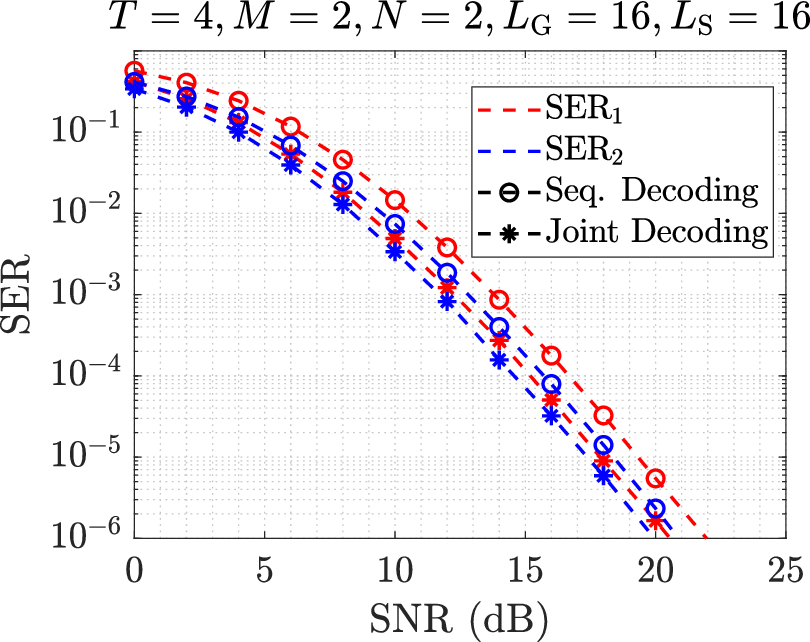}}
        \hfill
\subfloat[\normalsize $R_{\mathrm{G}}=0.75$,  $R_{\mathrm{S}}=1.25$ $(\textrm{bps/Hz})$  \label{fr1_3}]{%
      \includegraphics[width=0.33\linewidth]{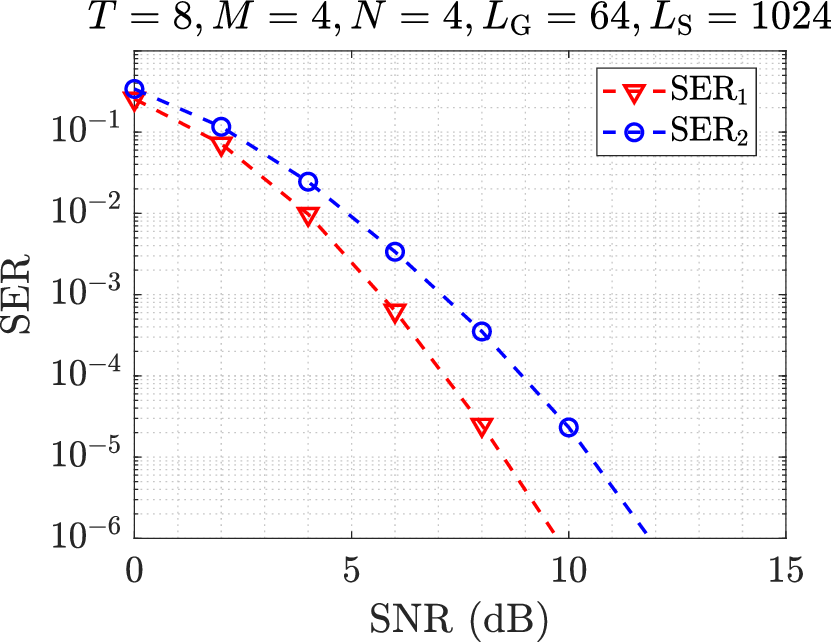}}
\caption{SER results for each layer for different values of $M$ with $T=2M$, $L_{\textrm{G}}$, and $L_{\textrm{S}}$, while maintaining a fixed total rate of $R=2$ bps/Hz. In (a) and (b), the performance of both the sequential and joint decoders is shown, whereas in (c), only the sequential decoder is evaluated.}
\label{fr1}
\end{figure*}

We first evaluate the SER performance of the two layers for different values of $M$ with $T=2M$, $N$, $L_{\textrm{G}}$, and $L_{\textrm{S}}$. For the second-layer decoding, perfect CSIR ($\hat{\mathbf{H}} = \mathbf{H}$) is assumed. This implies that $T_c \gg T$, so that approximately perfect CSIR can be obtained through the transmission of a sufficiently long known sequence without significantly affecting the overall rate. The case of imperfect CSIR is evaluated in Section \ref{TCNR}. Considering sequential decoding, the first-layer results are entirely independent of both the CSIR and the second-layer configuration, which constitutes a major advantage of the proposed design. Hence, the first layer can be evaluated separately, and its performance remains unchanged even when the second layer is deactivated. Consequently, these results coincide with those reported in \cite{gd}, since the same $\mathcal{U}$ constellations are considered and noncoherent detection is applied in all cases. Finally, in this case we assume no channel correlation in, with $\gamma_{\mathrm{t}}=\gamma_{\mathrm{r}}=0$.

Fig. \ref{fr1} shows the performance of each layer for different values of $M$, with $T = 2M$, $N$, $L_{\textrm{G}}$, and $L_{\textrm{S}}$, while maintaining a fixed total rate of $R = 2$ bps/Hz. For additional comparison, we also include the results obtained with joint decoding of both layers, as described in \eqref{mls_joint}, in Fig. \ref{fr1_1} and Fig. \ref{fr1_2}. Note that, for the configuration considered in Fig. \ref{fr1_3}, joint decoding at the selected rate would require comparisons over $2^{16}$ constellation points, making it impractical for implementation. In contrast, the sequential decoder requires only $2^6 + 2^{10}$ comparisons for that same configuration.

As expected, for the same total rate, the SER of both layers is enhanced by increasing the number of antennas. Furthermore, it can be observed that both layers exhibit comparable SER performance in Fig. \ref{fr1_1} and Fig. \ref{fr1_2} with $L_{\textrm{G}} = L_{\textrm{S}}$, which is desirable to prevent the overall BER (Section \ref{STBC}) from being dominated by the weaker layer. Moreover, the performance of the second layer is slightly better in those cases, even though its points are decoded after the first layer when sequential decoding is considered. This validates a key feature of the proposed constellation design described in Section \ref{SDCon}: the indices of different Stiefel representatives can still be distinguished even when the incorrect subspace is selected. As a result, sequential decoding shows only a slight performance degradation (below $2$ dB in the presented results) compared with joint decoding with CSIR, while its computational complexity scales with $L_\mathrm{S}+L_\mathrm{G}$ rather than $L_\mathrm{S}L_\mathrm{G}$ (Section \ref{CC}).

\begin{figure}
\centering
\includegraphics[width=0.7\linewidth]{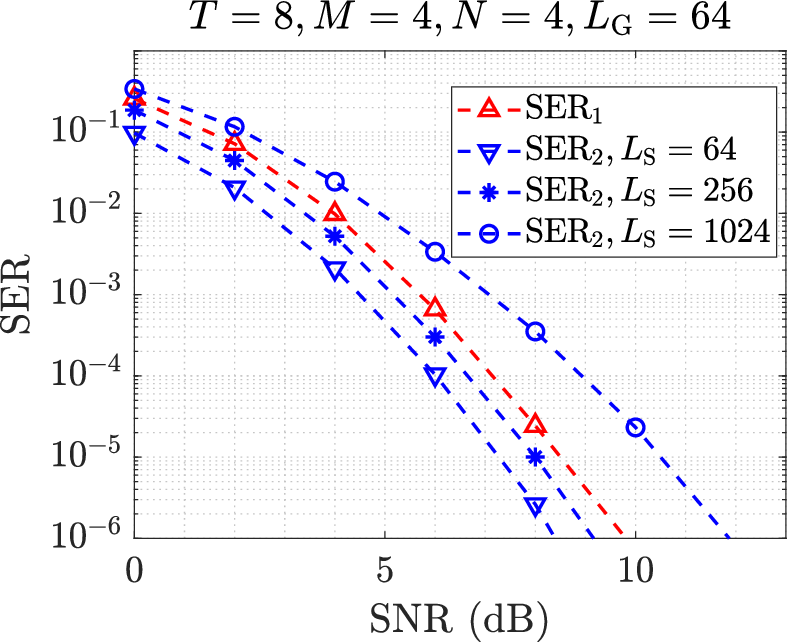}
\caption{SER results for different values of $L_{\textrm{S}}$ for $T=8$, $M=4$, $N=4$, and $L_{\textrm{G}}=64$.}
\label{fr2}
\end{figure}

In Fig. \ref{fr1_3}, since $L_{\textrm{G}}$ is restricted to $L_{\textrm{G}} \leq 4M^2=64$, we select $L_{\textrm{S}} = 1024$ to achieve $R = 2$ bps/Hz, with the disparity in the number of points resulting in $\mathrm{SER}_2 > \mathrm{SER}_1$. Nevertheless, even with $L_{\mathrm{S}} \gg L_{\mathrm{G}}$, the constellation design efficiently separates Stiefel points in the same equivalence class, keeping $\mathrm{SER}_2$ within approximately $2$ dB of $\mathrm{SER}_1$ at high SNR. To examine this behavior in greater detail, Fig. \ref{fr2} presents the $\mathrm{SER}_2$ results for different values of $L_{\mathrm{S}}$, with $T = 8$, $M = 4$, $N = 4$, and $L_{\mathrm{G}} = 64$. While $\mathrm{SER}_1$ is unaffected by $L_{\mathrm{S}}$, for $L_{\mathrm{S}} = L_{\mathrm{G}} = 64$ and $L_{\mathrm{S}} = 256$, we observe that $\mathrm{SER}_2 < \mathrm{SER}_1$. On the other hand, as expected, increasing $L_{\mathrm{S}}$ increases $R_{\mathrm{S}}$ but progressively degrades the error performance of the second layer.

\begin{figure}
\centering
\includegraphics[width=0.7\linewidth]{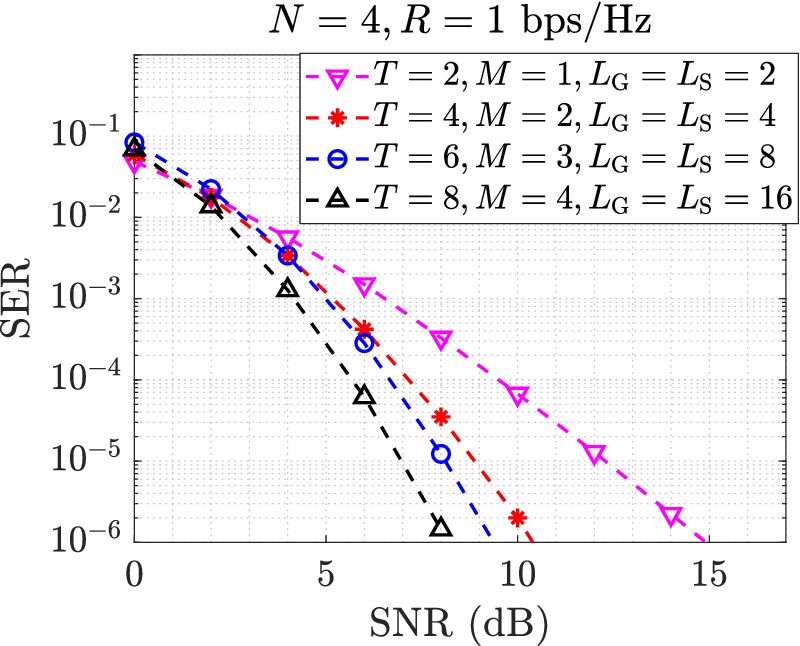}
\caption{$\mathrm{SER}$ results with  $L_{\mathrm{G}} = L_{\mathrm{S}}$ across different values of $T$ and $M$ for fixed values of receive antennas $N$ and spectral efficiency $R$.}
\label{fr3}
\end{figure}

Finally, to examine the impact of the number of transmit antennas, Fig. \ref{fr3} shows the overall system performance (Section \ref{SL}) for the case $L_{\mathrm{G}} = L_{\mathrm{S}}$, across different values of $T$ and $M$, for fixed values of the number of receive antennas $N=4$ and spectral efficiency $R=1$ bps/Hz. It is shown that the proposed constellation design is able to exploit the increased dimensionality of both the Grassmann and Stiefel manifolds provided by a larger number of antennas to improve error performance, while preserving the property that each space-time matrix has a single nonzero entry per row, regardless of the value of $M$. 

\subsection{Comparison with Other Constellation Designs}\label{COCD}

To compare the proposed constellation designs for both layers with other alternative under sequential decoding, we consider the case $T=4$, $M=N=2$, and $L_{\textrm{G}}=L_{\textrm{S}} = 16$. In this setting, the use of an Alamouti STBC for the alphabet $\mathcal{R}$ is of particular interest \cite{multicasting}:

\begin{equation}
\label{Rala}
    \mathcal{R}_{\textrm{Ala.}} = \frac{1}{\sqrt{2}}\begin{bmatrix}
q_1 & q_2\\
-q_2^* & q_1^*,
\end{bmatrix},
\end{equation} where $q_1$ and $q_2$ are complex symbols drawn from a $4$-QAM (or QPSK) constellation. For $M=2$ and $L_{\textrm{S}}=16$, this construction corresponds exactly to the $16$ vertices of a four-dimensional hypercube inscribed in the unit $3$-sphere $\mathbb{S}^3$. In this particular case, it provides a deterministic construction that achieves excellent performance in terms of DP, with a pairwise minimum of $\mathrm{DP}_{\mathrm{S}}=1$ in \eqref{dps}. Although this construction does not generalize with comparable performance to arbitrary values of $M$ and $L_{\textrm{S}}$, for $M=2$ and $L_{\textrm{S}}=16$ it provides a useful reference for evaluating the second layer.

\begin{figure}
\centering
\includegraphics[width=0.7\linewidth]{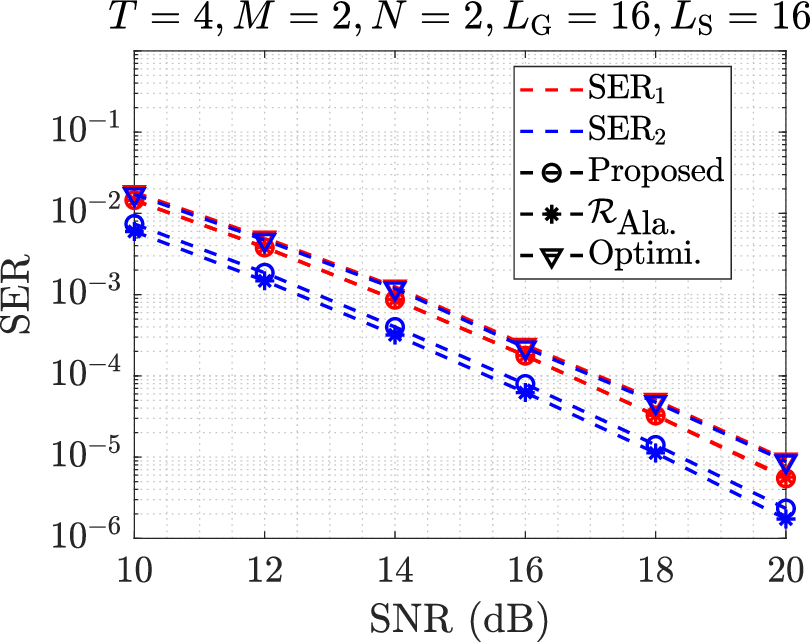}
\caption{$\mathrm{SER}$ results comparing three different constellations designs: the proposed design for both the first layer, $\mathcal{U}$, and the second layer, $\mathcal{R}$, (Proposed); the combination of the same $\mathcal{U}$ alphabet with  $\mathcal{R}_{\mathrm{Ala.}}$ in \eqref{Rala} ($\mathcal{R}_{\mathrm{Ala.}}$); and the combination of $\mathcal{R}_{\mathrm{Ala.}}$ with an optimization-based approach for obtaining $\mathcal{U}$ (Optimi.).}
\label{frcomp}
\end{figure}

Fig. \ref{frcomp} presents the $\mathrm{SER}$ results for three different constellation designs: the proposed design for both the first layer, $\mathcal{U}$, and the second layer, $\mathcal{R}$, (Proposed); the combination of the same $\mathcal{U}$ alphabet with  $\mathcal{R}_{\mathrm{Ala.}}$ in \eqref{Rala} ($\mathcal{R}_{\mathrm{Ala.}}$); and the combination of $\mathcal{R}_{\mathrm{Ala.}}$ with an optimization-based approach for obtaining $\mathcal{U}$, based on gradient-ascent of the DP in the Grassmann manifold \eqref{dp}, proposed in \cite{dp}, whose implementation is publicly available in \cite{thesis} (Optimi.). Remarkably, the proposed design achieves nearly the same $\mathrm{SER}_2$ performance as the optimal $\mathcal{R}_{\textrm{Ala.}}$ approach, while using only diagonal unitary matrices, thereby providing the associated benefits of sparsity and reduced computational complexity. On the other hand, although the matrices in $\mathcal{R}_{\textrm{Ala.}}$ are not diagonal (Section \ref{SDCon}), their combination with the structured design for $\mathcal{U}$ yields good performance under sequential decoding. This is not the case when $\mathcal{U}$ obtained through optimization (Optimi.), where the effect of $\mathrm{SER}_1$ limiting the performance of $\mathrm{SER}_2$ is clearly visible. A similar trend can be observed when comparing these results with those in \cite[Fig. 17]{multicasting}, noting that the SNR reference used in that work is shifted by $3$ dB with respect to ours. There, combining an optimization-based design for $\mathcal{U}$ with the same $\mathcal{R}_{\textrm{Ala.}}$ adopted here, also results in poor $\mathrm{SER}_2$ performance under sequential decoding. By contrast, when a structured first-layer design is used (in that work, the Exp-Map scheme \cite{expmap2}), $\mathrm{SER}_2$ can clearly outperform $\mathrm{SER}_1$. 

Considering the explicit conditions that favor sequential decoding identified in Section \ref{SDCon}, a similar trend is observed for the Exp-Map structured design \cite{expmap2} compared with the design of \cite{gd}: many (although not all) combinations of points in $\mathcal{U}$ and $\mathcal{R}$, with $\mathcal{R}$ either diagonal or given by \eqref{Rala} for $M=2$, satisfy \eqref{eq:proof}. This results in a favorable distribution of the resulting points in the Stiefel manifold in terms of Frobenius distance, yielding a significant improvement in $\mathrm{SER}_2$ performance compared with unstructured optimization-based designs, for which \eqref{eq:proof} is satisfied by considerably fewer point combinations. For the first layer, the constellation in \cite{gd} is adopted instead of that in \cite{expmap2} due to its sparsity, reduced computational complexity, and improved SER performance compared with other structured approaches, albeit with the limitation $L_{\mathrm{G}}\leq 4M^2$.

\subsection{Comparison with Classical STBC Schemes}\label{STBC}

\begin{figure*}
\centering
\subfloat[\normalsize $R_{\mathrm{G}}=1$,  $R_{\mathrm{S}}=1$ $(\textrm{bps/Hz})$ \label{fr4_1}]{%
      \includegraphics[width=0.33\linewidth]{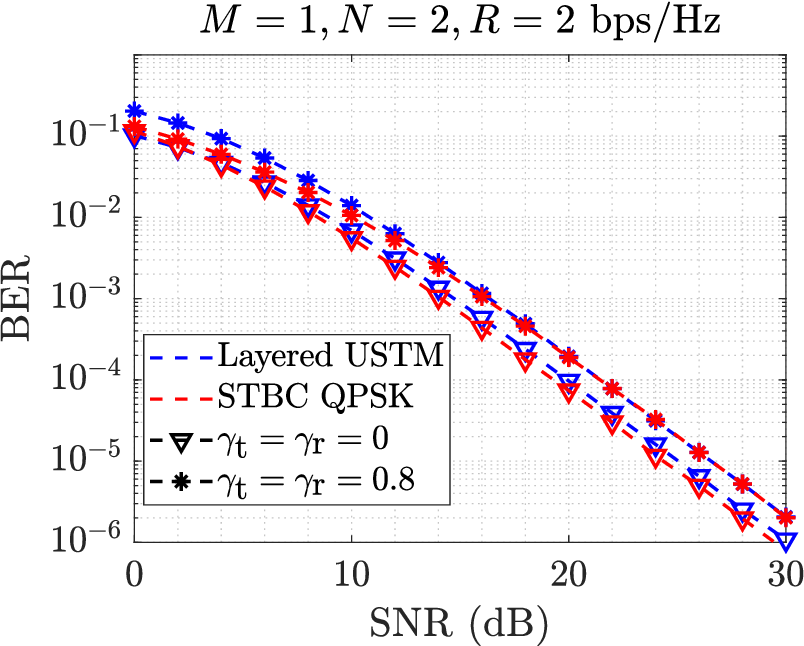}}
      \hfill
\subfloat[\normalsize  $R_{\mathrm{G}}=1$,  $R_{\mathrm{S}}=1$ $(\textrm{bps/Hz})$   \label{fr4_2}]{%
      \includegraphics[width=0.33\linewidth]{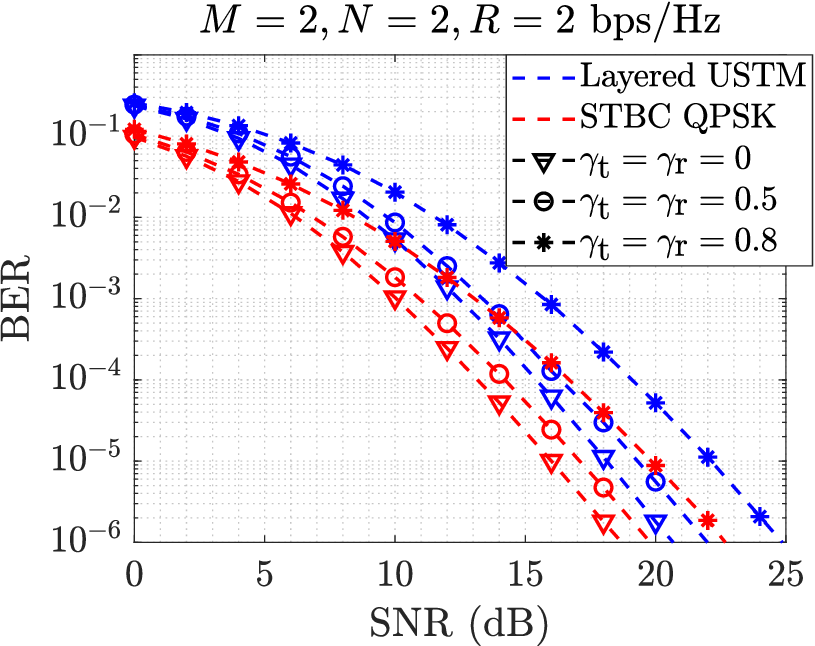}}
        \hfill
\subfloat[\normalsize $R_{\mathrm{G}}=0.75$,  $R_{\mathrm{S}}=0.75$ $(\textrm{bps/Hz})$  \label{fr4_3}]{%
      \includegraphics[width=0.33\linewidth]{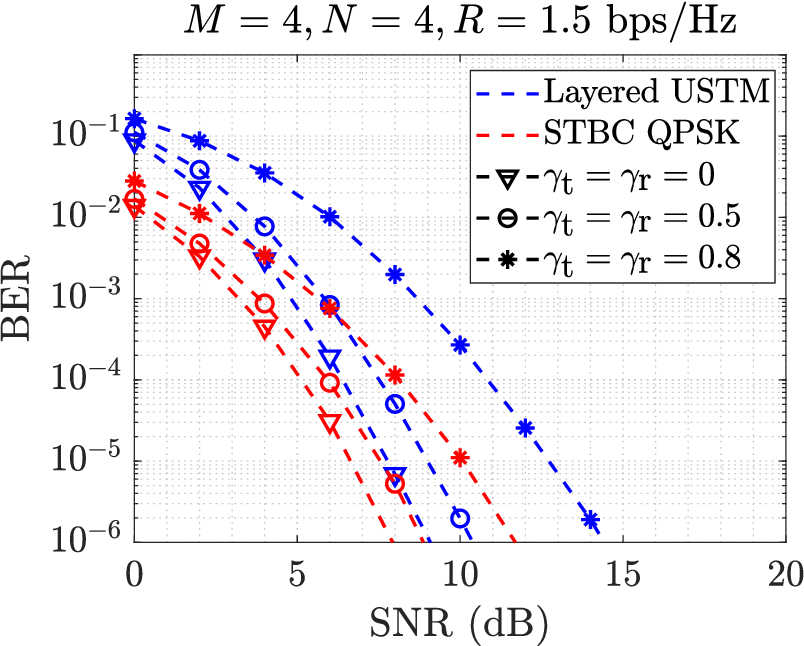}}
\caption{BER comparison between the proposed design (Layered USTM) and classical STBC schemes for different values of $M$, $N$, $R$, and channel correlation.}
\label{fr4}
\end{figure*}

We now compare the BER of the proposed USTM strategy against classical STBC designs that rely on CSIR but not on CSIT. Again, perfect CSIR is assumed in all cases, while different levels of channel correlation are now evaluated. Fig.~\ref{fr4} shows the comparison for different values of $M$, $N$, and $R$, with $R_{\mathrm{G}} = R_{\mathrm{L}}$ for the proposed USTM strategy. In all cases, the standard STBC strategies use a QPSK constellation and maximum-ratio combining at the receiver. For $M = 2$, the Alamouti scheme is used for transmission \cite{alamouti}. For $M = 4$, a $3/4$ rate scheme is selected to preserve full diversity and symbol orthogonality at the cost of transmitting only three symbols over four time slots \cite{book_stbc}, \cite{stbc34} (note that achieving full diversity, symbol orthogonality, and full rate simultaneously is impossible for $M > 2$ \cite{book_stbc}). This yields a rate of $R = 1.5$ bps/Hz, and the proposed USTM design is adjusted accordingly to ensure a fair comparison. Finally, note that the STBC schemes use $T = M$ rather than $T = 2M$, which is naturally accounted for when calculating their rate.

For $M = 1$ and $N=2$, the results are nearly identical for both strategies and remain almost unaffected even under strong correlation, $\gamma_{\mathrm{t}} = \gamma_{\mathrm{r}} = 0.8$, which is expected due to the reduced number of antennas. For $M=N=2$, the STBC Alamouti scheme show a consistent improvement of about 2 dB for BER values below $10^{-3}$, and both STBC QPSK and USTM degrade smoothly with channel correlation, being the gap below $1$ dB for moderate correlation ($\gamma_{\mathrm{t}}=\gamma_{\mathrm{r}}=0.5$), and up to $5$ dB for strong correlation ($\gamma_{\mathrm{t}}=\gamma_{\mathrm{r}}=0.8$). In the case of $M = N = 4$, the STBC QPSK strategy shows only an improvement of approximately $1$ dB over USTM for BER values below $10^{-4}$ in the absence of correlation. Both schemes respond gracefully to moderate correlation ($\gamma_{\mathrm{t}} = \gamma_{\mathrm{r}} = 0.5$), exhibiting a degradation of only about $1$ dB. Under strong correlation ($\gamma_{\mathrm{t}} = \gamma_{\mathrm{r}} = 0.8$), the STBC QPSK scheme degrades by up to $4$ dB, while the USTM scheme degrades by up to $5$ dB. Finally, it must be noted that the Alamouti scheme requires both antennas to be active during each time period, whereas the $3/4$ rate scheme requires three active antennas. On the other hand, the proposed USTM system requires only a single active antenna per symbol in the single-carrier case, potentially enabling implementation with a single radio-frequency (RF) chain and an antenna-switching mechanism \cite{gd}. In the case of multicarrier OFDM, this property translates into a single active antenna per subcarrier during the transmission of an OFDM symbol. Although this requires $M$ RF chains, this property may be useful for integrating communications with other functionalities, as envisioned in the ISAC framework. In this context, it is particularly relevant that previous works have identified USTM as a space-time structure of interest for ISAC from an information-theoretic perspective \cite{isac_info}.

\subsection{Impact of Coherence Time Interval}\label{TCNR}

We now evaluate perhaps the most important aspect of applying USTM when compared with other approaches, namely, how the system performs under short or moderate coherence intervals, where the transmission of pilot signals must be minimized or even removed. For the limit case of USTM, $T_c = 2M = T$, our system would choose to completely forgo the second layer and rely solely on noncoherent detection of the first layer. In contrast, for a classical STBC scheme relying on CSIR with $T = M$, we would have $T_c = 2M = 2T$ and would therefore spend half of the coherence interval transmitting orthogonal pilot signals and the other half transmitting data \cite{mimo_tran}. A comparison of both strategies under these conditions is presented in \cite[Fig. 9]{gd}, where it is shown that the noncoherent approach based on Grassmannian signaling surpasses the coherent approach due to the impact of imperfect CSIR in the coherent schemes. These results apply directly to our proposal, since the first-layer constellations are identical to those presented in \cite{gd}.

\begin{figure*}
\centering
\subfloat[\normalsize $L_{\mathrm{G}}=L_{\mathrm{S}}=4$,  $L_{\mathrm{N}}=16$ \label{fr5_1}]{%
      \includegraphics[width=0.33\linewidth]{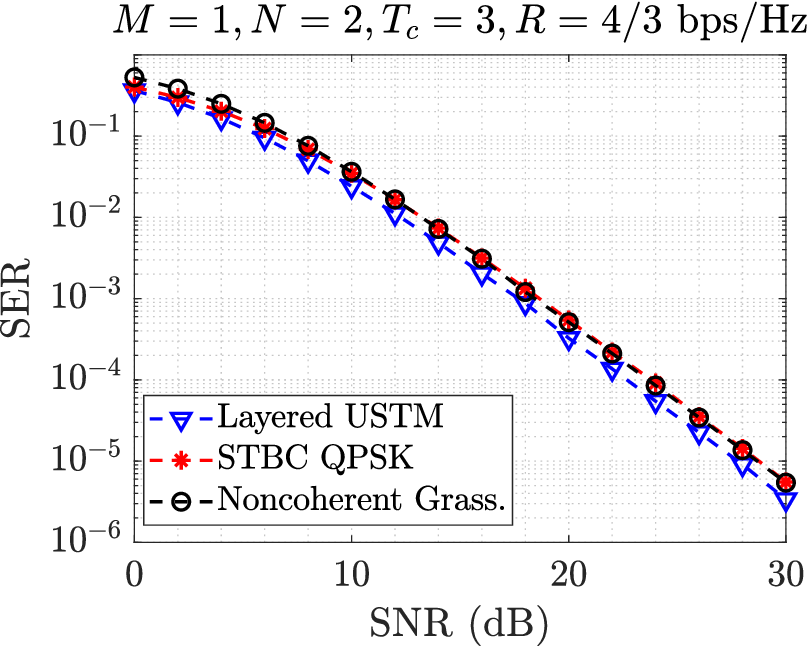}}
      \hfill
\subfloat[\normalsize  $L_{\mathrm{G}}=L_{\mathrm{S}}=16$,  $L_{\mathrm{N}}=256$  \label{fr5_2}]{%
      \includegraphics[width=0.33\linewidth]{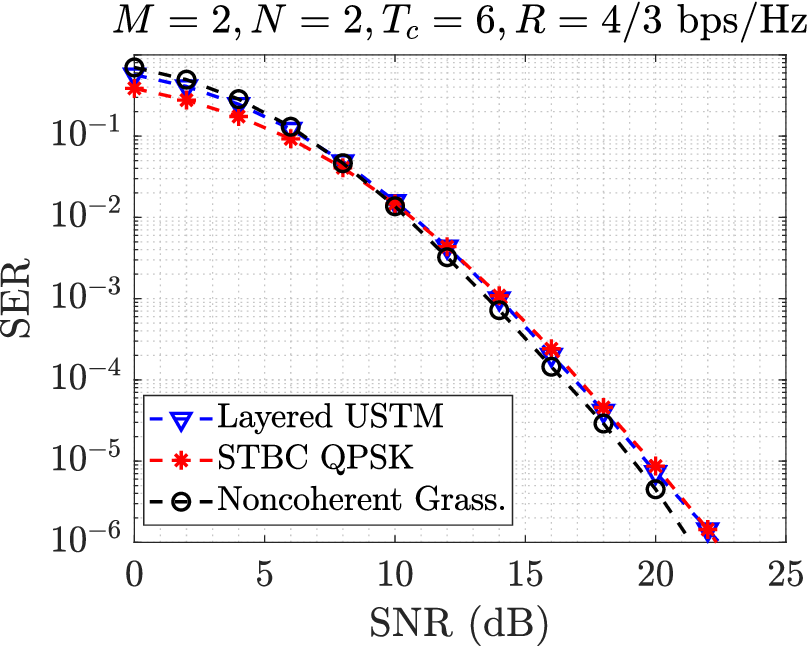}}
        \hfill
\subfloat[\normalsize $L_{\mathrm{G}}=L_{\mathrm{S}}=64$ \label{fr5_3}]{%
      \includegraphics[width=0.33\linewidth]{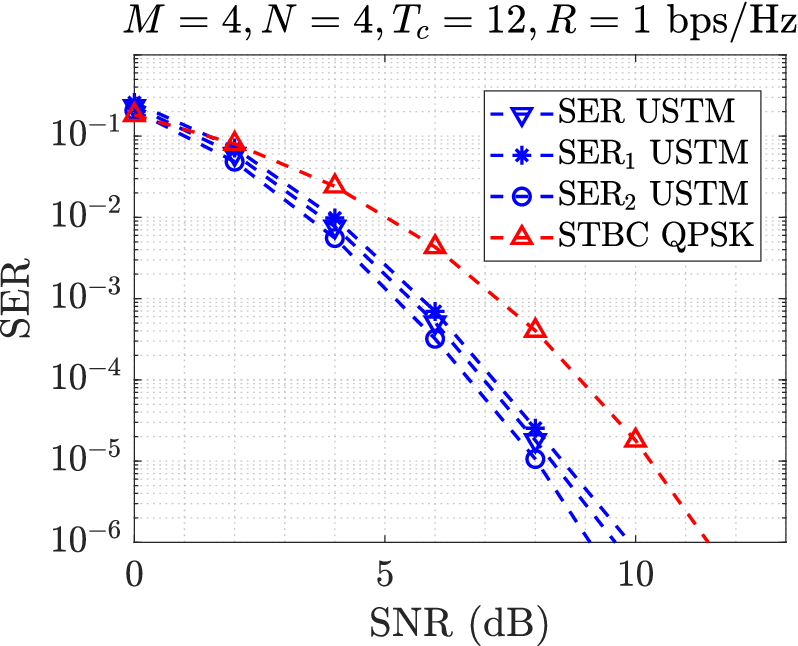}}
\caption{SER comparison between the proposed design, classical STBC, and noncoherent Grassmannian signaling for different values of $M$, $N$, $R$, and $T_c$.}
\label{fr5}
\end{figure*}

The case of larger (but still limited) values of $T_c$ can be clearly illustrated through an example. Consider $M = 2$ and $T_c = 6$. In this case, the proposed USTM scheme can use the first two time slots to transmit orthogonal pilot symbols and estimate the channel matrix, $\hat{\mathbf{H}}$, via maximum-likelihood pilot observation. However, due to AWGN and the reduced number of pilot symbols, perfect CSIR cannot be assumed, i.e., $\hat{\mathbf{H}} \neq \mathbf{H}$. The remaining four time intervals are then used to transmit a single Stiefel representative with $T = 2M = 4$. For a classical STBC scheme with $T = M$, the channel estimation procedure is identical, but two space-time matrices can be transmitted within the same coherence interval. In both cases, the effective transmission rate is scaled by a factor of $2/3$ due to pilot transmission during $T_c/3$ time intervals. Finally, there is a third possibility: applying noncoherent Grassmannian signaling with an increased block length $T$, setting $T_c = T$, and completely forgoing channel estimation. The comparison among these three strategies is presented in Fig.~\ref{fr5}, where analogous scenarios are considered for $M = 1$ and $M = 4$, and the number of points of the fully noncoherent strategy is denoted by $L_{\mathrm{N}} = 2^{RT_c}$. Since $L_{\mathrm{G}} = L_{\mathrm{S}}$ in the proposed design, we obtain $\mathrm{SER} = (\mathrm{SER}_1 + \mathrm{SER}_2)/2$ (Section \ref{SL}). For the Grassmannian constellation design with $T > 2M$, we use the DP gradient-ascent optimization proposed in \cite{dp}.

We observe that, for $M = 1$ and $N = 2$ with $T_c = 3$, all three schemes yield similar performance, with the proposed USTM design outperforming the others by approximately $1$ dB. For $M = N = 2$, a similar trend is observed, although in this case the noncoherent scheme provides the best performance, albeit only by about $1$ dB at high SNR. However, note that this approach requires $L_{\mathrm{N}} = 256$ points and therefore $256$ point comparisons for ML decoding, whereas the proposed design requires only $L_{\mathrm{G}} + L_{\mathrm{S}} = 32$ point comparisons. For the case $M = N = 4$ with $T_c = 12$ and $R = 1$ bps/Hz, the noncoherent approach would require the comparison of $2^{12} = 4096$ points (space-time matrices), which would make it unfeasible for most practical systems, and also would add significant complexity to the constellation design via numerical optimization. Therefore, this case is not plotted. Instead, we present the SER values of each layer, as well as the overall SER, for the proposed USTM strategy, which in this scenario achieves an improvement of approximately $2$ dB over the STBC QPSK scheme. The differences between Fig.\ref{fr4} and Fig.\ref{fr5} can be explained by the different impact of imperfect CSIR on both schemes. In the proposed USTM strategy, the first layer is completely unaffected, while the second layer experiences only a minor degradation (see Fig. \ref{fr2} versus Fig. \ref{fr5_3}). In contrast, conventional STBC schemes rely on CSIR not only for symbol estimation, but also for mitigating inter-stream interference. Consequently, imperfect CSIR introduces interference between different symbols, resulting in greater system degradation than in the proposed design.

Importantly, the observed BER and SER results can be related to previous information-theoretic conclusions. First, under the assumption of equal-energy signals, USTM remains an appealing approach regardless of whether CSI is available at the receiver \cite{gt2}, i.e., in both coherent and noncoherent scenarios. Second, as concluded in \cite{gt2_2}, USTM is attractive from both information-theoretic and SER perspectives, and for both uncorrelated and correlated channels when channel-correlation information is unavailable. Finally, as studied in \cite{gt3}, for limited coherence intervals, the insights on the suitable number of antennas for a pilot-based scheme are aligned with those for the noncoherent case, enabling an efficient combination of both architectures within a common hardware system, as proposed here.

\section{Conclusion}\label{Conclusion}

The proposed layered USTM constellation combines a previous Grassmannian design for the first noncoherent layer with a novel systematic design for the second layer that applies to an arbitrary number of antennas and $L_{\mathrm{S}}$ points. Importantly, under sequential decoding, the proposed design prevents the SER of the first layer from severely limiting that of the second layer, as observed with other designs. Furthermore, the proposed combination yields a sparse structure with a single nonzero element per row in the space-time matrices, significantly reducing decoding complexity. Finally, it enables a simple yet effective bit labeling.

Compared with noncoherent schemes based on Grassmannian signaling, the considered system provides the same features when only the first layer is activated. When the second layer is enabled, the performance of the first layer remains unchanged under sequential decoding, while coherent detection becomes possible using the same constellation structure as in the noncoherent case, with the additional information conveyed through variations among the Stiefel representatives. Focusing on coherent detection, the proposed layered USTM scheme achieves SER and BER performance comparable to, albeit slightly worse than, classical STBC schemes that rely on CSIR but not CSIT when perfect CSIR is assumed, with a gap below $2$ dB even relative to the Alamouti scheme for $M=2$, which provides the best performance relative to the number of antennas among the considered STBC schemes. The proposed USTM scheme also exhibits graceful performance degradation under channel correlation. On the other hand, it offers three important advantages. First, it provides the versatility of integrating noncoherent and coherent detection, enabling reliable operation across a wide range of wireless communication scenarios. Second, the proposed constellation provides sparse space-time matrices while preserving the fixed USTM structure. In single-carrier transmission, this property potentially enables implementation with a single transmit RF chain, reducing hardware cost and complexity. In OFDM, this translates into a single active antenna per subcarrier during each OFDM symbol, a feature that may be advantageous for integration within ISAC systems, for which previous studies have indicated the potential of the USTM structure. Third, and more importantly, for intermediate coherence intervals where perfect CSIR cannot be assumed, which constitute the most relevant application scenario for this system, the proposed layered USTM scheme outperforms conventional STBC schemes while, through sequential decoding, maintaining lower computational complexity compared with the application of purely Grassmannian signaling. 


\appendices
\section*{Appendix}

\subsection{Proof of Theorem~\ref{thm:UR}}
\label{app:proof}
Let $\mathbf{U} \coloneqq \mathbf{U}_g^H\mathbf{U}_{\tilde{g}}$ and $\mathbf{R} \coloneqq \mathbf{R}_{\tilde{s}}\mathbf{R}_s^H$. A straightforward computation shows that
\begin{equation}
\begin{split}
    ||\mathbf{U}_g\mathbf{R}_s-\mathbf{U}_{\tilde{g}}\mathbf{R}_{\tilde{s}}||_F^2 & = ||\mathbf{U}_g-\mathbf{U}_{\tilde{g}}\mathbf{R}||_F^2 \\
    & = 2M-2\mathfrak{Re}\left(\operatorname{tr}\left(\mathbf{U}\mathbf{R}\right)\right),
\end{split}
\end{equation}
where we have used $\mathbf{R} \in U(M)$ and $\mathbf{U}_g^H\mathbf{U}_g=\mathbf{U}_{\tilde{g}}^H\mathbf{U}_{\tilde{g}}=\mathbf{I}_M$. Note that $\mathbf{R} = \mathbf{I}_M$ when $s=\tilde{s}$, so \eqref{eq:proof} is equivalent to
\begin{equation}
\label{eq:proof2}
    \mathfrak{Re}\left(\operatorname{tr}\left(\mathbf{U}\right)\right) \geq \mathfrak{Re}\left(\operatorname{tr}\left(\mathbf{U}\mathbf{R}\right)\right).
\end{equation}

First, assume that $\mathbf{R}_{s}$ and $\mathbf{R}_{\tilde{s}}$ are diagonal, so that $\mathbf{R} = \operatorname{diag}(r_1,\ldots,r_M)$, with $|r_k|=1$. If the diagonal entries of $\mathbf{U}$, denoted $\mathbf{U}_{kk}$, are real and non-negative, then
\begin{equation}
\begin{split}
    \mathfrak{Re}\left(\operatorname{tr}\left(\mathbf{U}\right)\right) & = \sum_{k=1}^M \mathbf{U}_{kk} = \sum_{k=1}^M |\mathbf{U}_{kk}| \geq \sum_{k=1}^M \mathfrak{Re}\left(\mathbf{U}_{kk}r_k\right) \\ 
    & = \mathfrak{Re}\left(\operatorname{tr}\left(\mathbf{U}\mathbf{R}\right)\right),
\end{split}
\end{equation}
so \eqref{eq:proof2} holds for all diagonal $\mathbf{R}_{s},\mathbf{R}_{\tilde{s}} \in U(M)$.

Assume now arbitrary $\mathbf{R}_s, \mathbf{R}_{\tilde{s}} \in U(M)$, so that $\mathbf{R}$ can be any unitary matrix. Assume that $\mathbf{U}$ is hermitian and positive semi-definite, so that its singular values $\sigma_1,\ldots,\sigma_M$ coincide with its (non-negative) eigenvalues. Consider then the SVD $\mathbf{U}=\mathbf{Q}\mathbf{\Sigma}\mathbf{V}^H$, with $\mathbf{Q},\mathbf{V} \in U(M)$ and $\mathbf{\Sigma}=\operatorname{diag}(\sigma_1,\ldots,\sigma_M)$. Note that $\mathbf{A} \coloneqq \mathbf{V}^H\mathbf{R}\mathbf{Q} \in U(M)$, so
\begin{equation}
\begin{split}
    \mathfrak{Re}\left(\operatorname{tr}\left(\mathbf{U}\right)\right) & = \sum_{k=1}^M \sigma_k \geq \sum_{k=1}^M \sigma_k \mathfrak{Re}\left(\mathbf{A}_{kk}\right) \\
    & = \mathfrak{Re}\left(\operatorname{tr}\left(\mathbf{\Sigma}\mathbf{V}^H\mathbf{R}\mathbf{Q}\right)\right) = \mathfrak{Re}\left(\operatorname{tr}\left(\mathbf{U}\mathbf{R}\right)\right).
\end{split}
\end{equation}
Therefore, \eqref{eq:proof2} holds for all $\mathbf{R}_{s},\mathbf{R}_{\tilde{s}} \in U(M)$.

\bibliographystyle{ieeetr}
\bibliography{references}

\begin{thebibliography}{10}

\bibitem{csi1}
D.~J. Love, R.~W. Heath, V.~K. N.~Lau, D.~Gesbert, B.~D. Rao, and M.~Andrews, ``{An overview of limited feedback in wireless communication systems},'' {\em IEEE J. Sel. Areas Commun.}, vol.~26, pp.~1341--1365, Oct. 2008.

\bibitem{csi2}
O.~Elijah, C.~Y. Leow, T.~A. Rahman, S.~Nunoo, and S.~Z. Iliya, ``{A Comprehensive Survey of Pilot Contamination in Massive MIMO—5G System},'' {\em IEEE Commun. Surveys Tuts.}, vol.~18, pp.~905--923, Nov. 2016.

\bibitem{csit_imp}
R.~Chopra, C.~R. Murthy, and K.~Appaiah, ``{Adaptive Data-Aided Time-Varying Channel Tracking for Massive MIMO Systems},'' {\em IEEE Trans. Commun.}, vol.~72, pp.~5458--5472, Sept. 2024.

\bibitem{nonch_review}
K.-H. Ngo, D.~Cuevas, R.~D.~M. Gil, V.~M. Baeza, A.~G. Armada, and I.~Santamaria, ``{Noncoherent MIMO Communications: Theoretical Foundation, Design Approaches, and Future Challenges},'' {\em IEEE Open J. Veh. Technol.}, pp.~1--20, (early access) 2025.

\bibitem{diff_survey}
C.~Xu, N.~Ishikawa, R.~Rajashekar, S.~Sugiura, R.~G. Maunder, Z.~Wang, L.-L. Yang, and L.~Hanzo, ``{Sixty Years of Coherent Versus Non-Coherent Tradeoffs and the Road From 5G to Wireless Futures},'' {\em IEEE Access}, vol.~7, pp.~178246--178299, Dec. 2019.

\bibitem{index_mod}
M.~Vilà-Insa, A.~Martí, M.~Lamarca, and J.~Riba, ``{Low-Complexity Detection of Permutational Index Modulation for Noncoherent Communications},'' {\em IEEE Wireless Commun. Lett.}, vol.~14, pp.~3059--3063, July 2025.

\bibitem{energy_based}
M.~Vilà-Insa, A.~Martí, J.~Riba, and M.~Lamarca, ``{Quadratic Detection in Noncoherent Massive SIMO Systems Over Correlated Channels},'' {\em IEEE Trans. Wireless Commun.}, vol.~23, pp.~14259--14272, June 2024.

\bibitem{gt2}
B.~Hochwald and T.~Marzetta, ``{Unitary Space-Time Modulation for Multiple-Antenna Communications in Rayleigh Flat Fading},'' {\em IEEE Trans. Inf. Theory}, vol.~46, pp.~543--564, Mar. 2000.

\bibitem{gt3}
L.~Zheng and D.~Tse, ``{Communication on the Grassmann Manifold: a Geometric Approach to the Noncoherent Multiple-Antenna Channel},'' {\em IEEE Trans. Inf. Theory}, vol.~48, pp.~359--383, Feb. 2002.

\bibitem{gt1}
T.~Marzetta and B.~Hochwald, ``{Capacity of a Mobile Multiple-Antenna Communication Link in Rayleigh Flat Fading},'' {\em IEEE Trans. Inf. Theory}, vol.~45, pp.~139--157, Jan. 1999.

\bibitem{gt2_2}
M.~Brehler and M.~Varanasi, ``{Asymptotic Error Probability Analysis of Quadratic Receivers in Rayleigh-fading Channels with Applications to a Unified Analysis of Coherent and Noncoherent Space-Time Receivers},'' {\em IEEE Trans. Inf. Theory}, vol.~47, pp.~2383--2399, Sept. 2001.

\bibitem{expmap2}
I.~Kammoun, A.~M. Cipriano, and J.-C. Belfiore, ``{Non-Coherent Codes over the Grassmannian},'' {\em IEEE Trans. Wireless Commun.}, vol.~6, pp.~3657--3667, Oct. 2007.

\bibitem{cubesplit}
K.-H. Ngo, A.~Decurninge, M.~Guillaud, and S.~Yang, ``{Cube-Split: A Structured Grassmannian Constellation for Non-Coherent SIMO Communications},'' {\em IEEE Trans. Wireless Commun.}, vol.~19, pp.~1948--1964, Mar. 2020.

\bibitem{ub}
D.~Cuevas, J.~Álvarez Vizoso, C.~Beltrán, I.~Santamaria, V.~Tuček, and G.~Peters, ``{Union Bound Minimization Approach for Designing Grassmannian Constellations},'' {\em IEEE Trans. Commun.}, vol.~71, pp.~1940--1952, Apr. 2023.

\bibitem{book_stbc}
H.~Jafarkhani, {\em {Space-Time Coding: Theory and Practice}}.
\newblock USA: Cambridge University Press, 1st~ed., 2010.

\bibitem{alamouti}
S.~Alamouti, ``{A Simple Transmit Diversity Technique for Wireless Communications},'' {\em IEEE J. Sel. Areas Commun.}, vol.~16, pp.~1451--1458, Oct. 1998.

\bibitem{2non}
K.~M. Attiah, K.~G. Seddik, and R.~H. Gohary, ``{Hierarchical coherent and non-coherent communication},'' in {\em 2017 51st Asilomar Conference on Signals, Systems, and Computers}, (Pacific Grove, CA, USA), pp.~1242--1247, IEEE, 2017.

\bibitem{multicasting}
K.~G. Seddik, R.~H. Gohary, M.~T. Hussien, M.~Shaqfeh, H.~Alnuweiri, and H.~Yanikomeroglu, ``{Multi-Resolution Multicasting Over the Grassmann and Stiefel Manifolds},'' {\em IEEE Trans. Wireless Commun.}, vol.~16, pp.~5296--5310, May 2017.

\bibitem{multires2}
M.~A. ElMossallamy, K.~G. Seddik, and R.~H. Gohary, ``{Multi-Resolution Multicasting Using Grassmannian Codes and Space Shift Keying},'' {\em IEEE Trans. Veh. Technol}, vol.~68, pp.~988--992, Nov. 2019.

\bibitem{p1}
N.~Endo, H.~Iimori, C.~Pradhan, S.~Malomsoky, and N.~Ishikawa, ``{Boosting Spectral Efficiency With Data-Carrying Reference Signals on the Grassmann Manifold},'' {\em IEEE Trans. Wireless Commun.}, vol.~23, pp.~10137--10149, Mar. 2024.

\bibitem{p2}
Y.~Yu, G.~B. Giannakis, and N.~Jindal, ``{Information-Bearing Noncoherently Modulated Pilots for MIMO Training},'' {\em IEEE Trans. Inf. Theory}, vol.~53, pp.~1160--1168, Mar. 2007.

\bibitem{gd}
{\'A}.~Pendás-Recondo and E.~Pendás-Recondo, ``{A Structured Family of Grassmannian Constellations via Geodesic Mapping for MIMO Noncoherent Communications},'' {\em IEEE Trans. Wireless Commun.}, vol.~25, pp.~21089--21104, Aug. 2026.

\bibitem{sparse}
J.~Asano, Y.~Hama, H.~Iimori, C.~Pradhan, S.~Malomsoky, and N.~Ishikawa, ``{Sparse Grassmannian Design for Noncoherent Codes via Schubert Cell Decomposition},'' {\em IEEE-OJCOMs}, vol.~7, pp.~3559--3573, Apr. 2026.

\bibitem{ghb}
T.~Bendokat, R.~Zimmermann, and P.~Absil, ``{A Grassmann Manifold Handbook: Basic Geometry and Computational Aspects},'' {\em Adv. Comput. Math.}, vol.~50, no.~6, 2024.

\bibitem{geom_alg}
A.~Edelman, T.~Arias, and S.~Smith, ``{The Geometry of Algorithms with Orthogonality Constraints},'' {\em SIAM J. Matrix Anal. Appl.}, vol.~20, no.~2, pp.~303--353, 1998.

\bibitem{cgrass}
A.~Machado and I.~M.~C. Salavessa, ``{Grassman Manifolds as Subsets of Euclidean Spaces},'' in {\em Differential Geometry (Santiago de Compostela, 1984)}, vol.~131 of {\em Research Notes in Mathematics}, pp.~85--102.
\newblock Pitman, Boston, MA (1985).

\bibitem{do1}
R.~H. Gohary and T.~N. Davidson, ``{Noncoherent MIMO Communication: Grassmannian Constellations and Efficient Detection},'' {\em IEEE Trans. Inf. Theory}, vol.~55, pp.~1176--1205, Mar. 2009.

\bibitem{grassma}
J.~Álvarez Vizoso, D.~Cuevas, C.~Beltrán, I.~Santamaria, V.~Tuček, and G.~Peters, ``{Constrained Riemannian Noncoherent Constellations for the MIMO Multiple Access Channel},'' {\em IEEE Trans. Inf. Theory}, vol.~69, pp.~4559--4578, July 2023.

\bibitem{do2}
M.~Beko, J.~Xavier, and V.~A.~N. Barroso, ``{Noncoherent Communication in Multiple-Antenna Systems: Receiver Design and Codebook Construction},'' {\em IEEE Trans. Signal Process}, vol.~55, pp.~5703--5715, Dec. 2007.

\bibitem{do3}
G.~Han and J.~Rosenthal, ``{Geometrical and Numerical Design of Structured Unitary Space–Time Constellations},'' {\em IEEE Trans. Inf. Theory}, vol.~52, pp.~3722--3735, Aug. 2006.

\bibitem{ubp}
M.~McCloud, M.~Brehler, and M.~Varanasi, ``{Signal Design and Convolutional Coding for Noncoherent Space-Time Communication on the Block-Rayleigh-Fading Channel},'' {\em IEEE Trans. Inf. Theory}, vol.~48, pp.~1186--1194, May 2002.

\bibitem{permu}
R.~L. Rivest, ``{Permutation Polynomials Modulo 2w},'' {\em Finite Fields and Their Applications}, vol.~7, pp.~287--292, Apr. 2001.

\bibitem{kron}
S.~Loyka, ``{Channel Capacity of MIMO Architecture using the Exponential Correlation Matrix},'' {\em IEEE Commun. Lett.}, vol.~5, pp.~369--371, Sept. 2001.

\bibitem{dp}
J.~Álvarez Vizoso, D.~Cuevas, C.~Beltrán, I.~Santamaria, V.~Tuček, and G.~Peters, ``{Coherence-Based Subspace Packings for MIMO Noncoherent Communications},'' in {\em 2022 30th European Signal Processing Conference (EUSIPCO)}, (Belgrade, Serbia), pp.~1906--1910, IEEE, 2022.

\bibitem{thesis}
D.~Cuevas, {\em {Advanced {Grassmannian} Constellation Designs for Noncoherent {MIMO} Communications}}.
\newblock PhD thesis, Universidad de Cantabria, Spain, Nov. 2024.
\newblock Software available at \url{https://github.com/diegocuevasfdez/grassbox/}.

\bibitem{stbc34}
V.~Tarokh, H.~Jafarkhani, and A.~Calderbank, ``{Space-Time Block Codes from Orthogonal Designs},'' {\em IEEE Trans. Inf. Theory}, vol.~45, pp.~1456--1467, July 1999.

\bibitem{isac_info}
Y.~Xiong, F.~Liu, Y.~Cui, W.~Yuan, T.~X. Han, and G.~Caire, ``{On the Fundamental Tradeoff of Integrated Sensing and Communications Under Gaussian Channels},'' {\em IEEE Trans. Inf. Theory}, vol.~69, pp.~5723--5751, Sept. 2023.

\bibitem{mimo_tran}
B.~Hassibi and B.~Hochwald, ``{How Much Training is Needed in Multiple-Atenna Wireless Links?},'' {\em IEEE Trans. Inf. Theory}, vol.~49, pp.~951--963, Apr. 2003.

\end{thebibliography}

\end{document}